\documentclass{article}

\usepackage{PRIMEarxiv}

\usepackage[utf8]{inputenc} 
\usepackage[T1]{fontenc}    
\usepackage{hyperref}       
\usepackage{url}            
\usepackage{booktabs}       
\usepackage{amsfonts}       
\usepackage{nicefrac}       
\usepackage{microtype}      
\usepackage{lipsum}
\usepackage{fancyhdr}       
\usepackage{graphicx}       
\usepackage{dcolumn}
\usepackage{mathtools}
\usepackage{lipsum}
\usepackage{amsmath,amssymb}
\usepackage{bm}
\graphicspath{{media/}}     

\title{Improved Radiative Transfer Corrections in Helium Emission Lines
}

\author{
  Oleg Kurichin, Alexandre Ivanchik \\
  Ioffe Institute \\
  St.Petersburg, Russia\\
  \texttt{\{O.A. Kurichin\} o.chinkuir@gmail.com} \\
}

\begin{document}
\maketitle

\begin{abstract}
We present a new detailed model of the He\,I collisional–recombination spectrum based on the most up-to-date atomic data. The model accounts for radiative transfer effects and the influence of a non-zero optical depth in He\,I lines arising from transitions to the metastable 2$^3$S state. The model reveals substantial deviations in the emissivities of the $\lambda3889$ and $\lambda7065$ lines in case of a non-zero optical depth, with previous models systematically underestimating and overestimating them by 5–20\%, respectively. In the optically-thin case, however, our results show good agreement with previous studies. Using the new model, we compute optically-thin emissivities for a wide set of UV, optical, and IR He\,I recombination lines over a fine grid of electron densities and temperatures which are typical for H II regions and planetary nebulae ($1 \leq n_e \leq 10^4$ cm$^{-3}$, $8000 \leq T_e \leq 22000$ K). In addition, we present new fitting formulae for radiative transfer corrections for several He\,I lines relevant to optical and near-infrared observations, covering $0 \leq \tau_{3889} \leq 10$ within the same density and temperature ranges. The accuracy of the obtained approximations is $\lesssim 0.1\%$ within the specified parameter range. These results can be readily implemented in modern codes for determining the primordial $^4$He abundance and are also applicable to a broader range of spectroscopic analyses of He\,I emission lines.
\end{abstract}


\section{\label{sec:intro}Introduction}

Analysis of helium emission lines is a powerful diagnostic tool for probing the physical conditions and chemical composition of a various astrophysical objects. In particular, modeling the emission spectra of low-metallicity blue compact dwarf galaxies has become the most reliable method for obtaining independent estimates of the abundance of primordial helium, Y$_p$. The combination of high-quality spectroscopic data and advanced photoionization models, that take into account various systematic effects, has made it possible to determine Y$_p$ with remarkable accuracy ($\lesssim 1.5\%$) \cite{he_1, he_2, he_3, he_4, he_5, he_6, he_7, he_8, he_9, he_10, he_11, he_12}. Consequently, one of the major directions of studies is the improvement of methods for quantifying and correcting for systematic uncertainties in photoionization models.

Among the most important sources of systematic uncertainty are the theoretical emissivities of He\,I lines and the treatment of radiative transfer effects. For many of the strongest transitions associated with the metastable $2^3$S level of He\,I, the optical depth is non-negligible. For example, observational estimates of optical depth in He\,I $\lambda3889$ line (it corresponds to transition $3^3P \rightarrow 2^3S$ and is one of the strongest He\,I emission lines in optical range) fall in the range $0 \lesssim \tau_{3889} \lesssim 5$ \cite{he_2, he_8, he_9}. The presence of a non-zero optical depth leads to the attenuation of some lines and the enhancement of others. Typically, line fluxes vary within $\sim 1-5\%$, but for some lines (like, e.g., $\lambda3889$ or $\lambda7065$) changes can reach values of $\sim 30\%$ at certain values of optical depth \cite{bss2002}. If these effects are not properly accounted for, they can introduce systematic shifts in the derived helium abundances, and consequently bias the estimates of Y$_p$.

The non-zero optical depth in some helium lines appears for the following reason. Unlike hydrogen, helium has two separate transition ladders: singlet ladder with total spin of electrons $S=0$ and triplet ladder with $S=1$. The transition from the first exited state $2^3S$ in triplet ladder to the ground state $1^1S$ (which is singlet) is strongly forbidden with $A = 1.27 \times 10^{-4}$ sec$^{-1}$ \cite{lp2001}, making the $2^3S$ level metastable. As a result, photons emitted from $n^3P \rightarrow 2^3S$ radiative transitions can be efficiently reabsorbed, significantly altering the populations of the affected levels. Through collisional mixing, these changes propagate to many other levels, ultimately leading to substantial deviations in the observed fluxes of several He\,I lines. To account for this effect, a so-called optical depth function, $f_\tau$, is included in photoionization models. The optical depth function is defined as the ratio between the emissivity calculated at a given optical depth $\tau$ and the emissivity in the optically thin case (e.g. \cite{bss2002}):
\begin{equation}
    f_\tau(\lambda, n_e, T) = \frac{E_\lambda(\tau\neq0, n_e, T)}{E_\lambda(\tau=0, n_e, T)}.
\end{equation}

In modern photoionization models used for primordial helium studies $f_\tau$ is implemented as a multiplicative correction factor  (see, e.g., \cite{he_2, he_6, he_8, he_9}). All modern models adopt the analytic function proposed in \cite{bss2002}:
\begin{equation}
    f_\tau(\lambda, n_e, T) = 1 + \frac{\tau}{2} \left( a_\lambda + \frac{T}{10^4~\text{K}} \times \sum_{i=0}^2 b^{(i)}_\lambda \,n_e^i \right)
    \label{eq:odf_bss}
\end{equation}
where $\tau$ is the optical depth in $\lambda3889$ line, and $a_\lambda, b_\lambda^{(i)}$ are fitting coefficients. Radiative transfer corrections for different lines are calculated using optical depths normalized to optical depth of this $\lambda 3889$ line\footnote{ He\,I $\lambda3889$ line (transition $3^3P \rightarrow 2^3S$) is used for normalization since it is the strongest line with a finite opacity in visible range. The closest alternatives to $\lambda3889$ such as $\lambda$3188 and $\lambda$10830 fall in the UV and IR ranges, respectively, and therefore are less suitable for the analysis and modeling of optical spectra.}.

Despite the wide use of function (\ref{eq:odf_bss}), it  suffers from several limitations.
\begin{itemize}
    \item Function (\ref{eq:odf_bss}) it assumes a linear dependence on $\tau$, which does not always hold for different transitions. For example, emissivity of $\lambda7065$ line corresponding to $3^3S \rightarrow 2^3P$ transition shows strong non-linear dependence on $\tau$ (see  \cite{bss2002}) which is not accounted for in function (\ref{eq:odf_bss}).
    \item The formal applicability range of function (\ref{eq:odf_bss}) set in \cite{bss2002} is limited to $1 \leq n_e \leq 300$ cm$^{-3}$, $12000 \leq T \leq 20000$ K, and $\tau \leq 2$. These constraints may be incompatible with the actual physical conditions in low-metallicity dwarf galaxies. For example, the object SBS 1135+581 studied in \cite{he_2} has the estimated temperature of $T = 11226 \pm 550$ K, and the object SBS 0335-052 has $T = 21780 \pm 3360$ K. Objects J0519+0007 and SBS 1152+579 from the same paper have estimated electron densities of $675\pm143$ cm$^{-3}$ and $452\pm197$ cm$^{-3}$ respectively. Objects SBS 0335--052, J0519+0007 and Mrk 450 all have estimated optical depth $\tau > 2$. These examples clearly demonstrate that the physical conditions in real galaxies may lie outside the applicability range of function (\ref{eq:odf_bss}). Therefore, its usage ``as it is'' for analysis of a large set of different galaxies with different physical conditions may introduce a systematic bias into the results.
    \item Function (\ref{eq:odf_bss}) was obtained as a fit of emissivities computed with the simplified models of He\,I recombination spectrum of \cite{bss1999, bss2002}, which rely on atomic data, largely computed in the late 1980s and early 1990s. The models are based on radiative data and recombination cascades from \cite{smiths1991, smiths1996}, combined with collisional data from \cite{sawey1993}. The only difference between the models of \cite{bss1999} and \cite{bss2002} is the number of modeled levels: $n_{\max} = 5$ for \cite{bss1999} and $n_{\max} = 20$ for \cite{bss2002}. In contrast, modern models of He\,I emission make use of much more accurate atomic data for all relevant processes and explicitly include a substantially larger number of $LS$-resolved levels. For instance, the model of \cite{porter2012}, which underlies essentially all recent studies of Primordial Helium, treats levels up to $n_{\max} = 100$ and includes an additional ``top-off'' pseudo-level to account for recombination into $n>100$ levels. Similarly, the recent model of \cite{delzanna2022} resolves $LS$-levels up to $n=40$ and adds $n$-resolved levels on top up to $n=100$. Detailed comparisons of the impact of updated atomic data on theoretical emissivities were carried out in \cite{porter2005, porter2012, delzanna2022}, demonstrating systematic discrepancies of order\,$\sim 1-2$\,\% between modern and older results for all key emission lines.
\end{itemize}

The listed shortcomings of the method from \cite{bss2002} for correction for radiative transfer in helium lines come from its internal properties. However, there is also an external problem with this method, related to its use in modern analysis. The direct use of function (\ref{eq:odf_bss}) in its current form may lead to uncontrollable systematic shifts in helium abundance estimates. The reason for that is the unintentional mixing of different atomic data when modeling the observed fluxes of helium lines. In modern photoionization codes, the helium-to-H$\beta$ line flux ratio is defined via the equation:
\begin{equation}
\frac{F(\lambda)}{F(\rm{H}\beta)} = y^+ \times \frac{E(\lambda, n_e, T)}{E(\rm{H}\beta, n_e, T)} \times f_\tau(\lambda, n_e, T) \times [\dots]
\label{eq:mix_prob}
\end{equation}

Here $F$ is the line flux (helium or H$\beta$), $y^+$ is the helium-to-hydrogen density ratio, $E$ is the theoretical emissivity at given electron density $n_e$ and temperature $T$, and $f_\tau$ is the optical depth function. The square bracket\,$[\dots]$ denotes additional well-known systematic corrections such as corrections for underlying stellar absorption, interstellar reddening etc (a detailed discussion can be found in, e.g., \cite{he_8}).

Ideally, the product $E(\lambda, n_e, T) \times f_\tau(\lambda, n_e, T)$ should represent the emissivity of line $\lambda$ at the chosen optical depth $\tau$, provided that both terms are derived from the same model of He\,I spectrum. In practice, however, $E(\lambda, n_e, T)$ is calculated using emission model based on modern atomic data \cite{porter2012, aver2013}, while $f_\tau$ relies on an outdated one \cite{bss1999}. The systematic biases introduced by combining emissivities from inconsistent spectrum models have never been quantitatively assessed.

In the absence of a better alternative, function (\ref{eq:odf_bss}) remains the universal choice in current Primordial Helium studies. The only recent attempt to overcome its restrictions is presented in study \cite{berg2025}, where authors used the original code of \cite{bss2002} to compute $f_\tau$ over a much broader 3D grid: $\log(n_e/\text{cm}^{-3}) = 0 - 6$ in 0.25 dex steps, $T = 5000 - 20000$ K in 500 K steps, and $\tau_{3889} = 0 - 15$ in steps of 1. Interpolation across this grid provides values of $f_\tau$ beyond the original fitting range. While this approach extends the practical use of the function (\ref{eq:odf_bss}), it does not address the more fundamental issue of the model of \cite{bss2002} -- the reliance on outdated atomic data.

The present study is aimed at developing a new model of the He\,I recombination spectrum based on the most up-to-date atomic data. Our primary goal is to construct this model and compute a refined grid of emissivities tailored to Primordial Helium studies. Equally important, we calculate a new optical depth function that resolves the shortcomings of the currently adopted prescription and, for the first time, ensures full consistency with the new set of helium emissivities. The key studies with which we compare our results are \cite{bss1999}, \cite{porter2012}, and \cite{delzanna2022}. For the optical depth function, we compare our results with \cite{bss2002}.

The collisional-recombination model of He\,I emissivities presented in this paper represents a global overhaul of the framework by \cite{bss1999, bss2002}, in which all relevant atomic data have been replaced with modern values and the number of modeled levels has been extended up to $n_{\max} = 50$. The key distinction of our model compared to more recent models of \cite{porter2012} and \cite{delzanna2022} lies in the treatment of collisional transitions between the low-lying excited states. In \cite{porter2012} and \cite{delzanna2022}, the corresponding rates were obtained by linear interpolation over temperature grids of average collision strengths from \cite{sawey1993} and \cite{bray2000} for transitions between levels with $n \leq 4$. In our model, the average collision strengths are instead calculated through direct integration of the high-resolution collision strengths data at a given temperature for transitions between levels with $n \leq 5$. Furthermore, we take into account the temperature dependence of the scaling factor used to calculate recombination rates onto Rydberg states with $l \leq 2$, which has not been considered previously. Altogether, our new model provides a more consistent and accurate description of the He\,I recombination spectrum, offering a solid foundation for future studies.

The structure of this paper is as follows. Section \ref{sec:at_data} is entirely devoted to the atomic data employed in our emissivity model. Section \ref{sec:he_model} describes the construction of the model itself. In Section \ref{sec:results}, we present the results of our emissivity calculations, provide an updated optical depth function, and compare our findings with those of the previous studies. Finally, our conclusions are summarized in Section \ref{sec:conclusions}.

\section{\label{sec:at_data}Atomic data}
We use the following atomic data to construct collisional-recombination model of Helium emissivities:
\begin{enumerate}
    \item Energy of $LS$-resolved sublevels $E_k$
    \item Einstein coefficients for transitions between sublevels $A_{ik}$
    \item Radiative recombination rates $\alpha_k$
    \item Collision (de-)excitation rates $q_{ik}$
    \item Rates of collisional ionization $C^{(ion)}_k$
\end{enumerate}

The main problem with the data is that for highly excited $LS$-resolved states in Helium there is no direct atomic data available. In subsections below we provide all necessary information on how different atomic data is handled. 

In the model we consider only single electron transitions between excited levels with the second electron remaining in the ground state. Hereafter we will use the following notation: by word ``level'' we mean the set of all sublevels with the same value of the principal quantum number $n$, and word ``sublevel'' or ``state'' stands for an $LS$-resolved state within a level. Throughout the paper we characterize an excited sublevel via the notation $\gamma_k = [n_k, l_k, s_k]$, where $n_k$ is the principal quantum number, $l_k$ us the azimuthal quantum number and $s_k = 2S_k + 1$ is the multiplicity of the sublevel $k$, which for Helium can be either 1 for singlet ladder or 3 for triplet ladder. Also we use notation $\gamma_u$ and $\gamma_l$ or just indices $u$ and $l$ to distinguish upper and lower sublevels in a transition respectively. The values of all fundamental constants used throughout this paper are taken from the PDG 2024 review \cite{pdg2024}.

\subsection{\label{subsec:energy}Energy levels}

The extensive table of energies for $LS$-resolved sublevels in Helium atom for $n\leq 10$ and $l\leq7$ for both multiplicity ladders are presented in \cite{dm2007}. We use these essentially exact values for lower lying excited sublevels. For sublevels with $n>10$ and/or $l>7$ we calculate energies from the Ritz quantum defect expansion as described in \cite{drake_handbook}. Energies of Rydberg states are expressed via the equation:
\begin{equation}
    E_k = \frac{R_M}{\nu^2} + \Delta E_k
\end{equation}

Here $R_M$ is mass-corrected Rydberg constant, $\nu$ is effective quantum number of a sublevel, the term $\Delta E_k$ is used to account for relativistic and mass polarization corrections in Helium. The explicit form of the latter is given via eq. (11.54) in \cite{drake_handbook}. The effective quantum number of a sublevel is calculated by iterative solution of equation
\begin{equation}
    \nu = n - \delta(\nu)
\end{equation}
Here $\delta(\nu)$ is the quantum defect defined by Ritz expansion \cite{drake_handbook, drake1991}:
\begin{equation}
    \delta(\nu) = \delta_0 + \frac{\delta_2}{(n - \delta)^2} + \frac{\delta_4}{(n-\delta)^4} + \dots
\end{equation}
Coefficients $\delta_i$ are constant values for a specified quantum numbers of $l$ and $s$. For our model we use coefficients from tab. 11.9 in \cite{drake_handbook}. A detailed discussion of how these coefficients are calculated and why such expansion is applicable to Rydberg states in Helium atom can be found in \cite{drake1991}. It should be noted, that in this paper we use data from year 2006 edition of Springer Handbook of Atomic, Molecular, and Optical Physics \cite{drake_handbook}, while there are different editions from 1996, 2006 and 2023. One can use either of these books since all data tables and accompanying formulas and explanations are the same throughout all editions with only difference being the number of the Chapter.

\subsection{\label{subsec:a_vals}A-values}

Einstein coefficients $A_{ul}$ for allowed radiative transitions between excited states in Helium atom are taken from \cite{dm2007}. As with the energy levels, in paper \cite{dm2007} only transitions between sublevels with $n\leq10$ and $l\leq7$ are considered. A-values for transitions involving sublevels with $n>10$ and/or $l>7$ are handled via one of the following methods.

\begin{figure}[ht]
    \centering
    \includegraphics[width=0.83\linewidth]{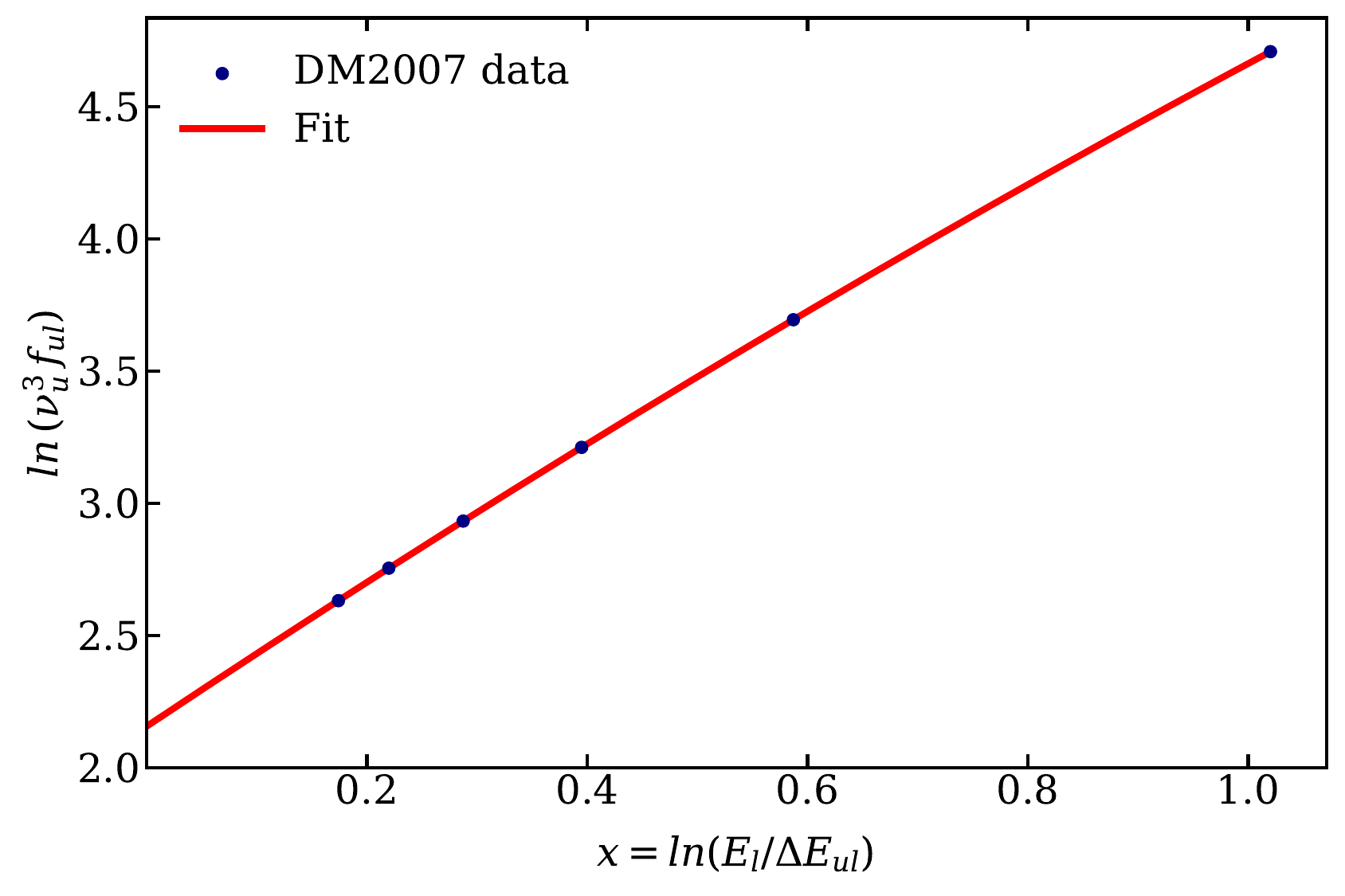}
    \caption{Extrapolation of oscillator strength data from \cite{dm2007} using $n^1F \rightarrow 4^1D$ transition as an example. Lower values of $x$ correspond to higher excited states. Abbreviation DM2007 represents paper \cite{dm2007}.}
    \label{fig:dm_extrapolation}
\end{figure}

If a transition $\gamma_u \rightarrow \gamma_l$ occurs between sublevels with $\gamma_u = [n_u>10,\, l_u\leq7,\, s_u]$ and $\gamma_l=[n_l\leq7,\,l_l\leq6,\, s_l]$ we calculate emission oscillator strength use extrapolation of data from \cite{dm2007} using series suggested in \cite{hs1998}:
\begin{equation}
    \ln(\nu^3_u\,f_{ul}) = a + bx + cx^2
\end{equation}
Here $\nu_u$ is effective quantum number of the upper sublevel, $x = \ln(E_l / \Delta E_{ul})$, $\Delta E_{ul} = E_u - E_l$. Effective quantum number and energies are calculated using method described in subsec. \ref{subsec:energy}. Fit parameters $a, b, c$ are calculated using trust region reflective algorithm based on data from \cite{dm2007}. Fit accuracy is better than $0.1\%$ for the majority of transitions under consideration. Example of such extrapolation for the $n^1F \rightarrow 4^1D$ transition is presented on Fig. \ref{fig:dm_extrapolation}. The values of fitting coefficients for all transitions under consideration are presented in Appendix \ref{app:a_val_coef}.

If a transition $\gamma_u \rightarrow \gamma_l$ occurs between sublevels with $\gamma_u = [n_u>10,\, l_u\leq7,\, s_u]$ and $\gamma_l=[n_l>10,\,l_l\leq7,\, s_l]$ emission oscillator strength are calculated using Coulomb approximation. Coulomb approximation is a reliable and simple method to calculate transition probabilities in multi electron atomic systems. It was firstly proposed in \cite{bates1949}. The method assumes that the initial and final states of the atom differ primarily in the wave function of the transiting electron, which is approximated by a Coulomb wave function with correct asymptotic behavior at large radii. In multi electron atoms wave functions of initial and final states are often expressed in terms of Whittaker functions, which are then integrated to obtain transition probability. In-depth review and discussion of the Coulomb approximation method can be found in e.g. \cite{hey2017}. 

In this paper we calculate oscillator strength for a transition $\gamma_u \rightarrow \gamma_l$ via the equation:
\begin{equation}
    f_{ul} = \frac{1}{(2L_l + 1)(2S_u+1)} \sum_{J_u, J_l} (2J_l + 1) \frac{\tilde{\Delta E_{ul}}}{|E_{J_u} - E_{J_l}|} f_{J_u J_l}
    \label{eq:multiplet_av}
\end{equation}

Here the summation is carried over all possible total angular momenta within a multiplet with energy weights, where $\tilde{\Delta E_{ul}}$ represents average transition energy. The oscillator strength for the individual members of the multiplet is obtained from the corresponding line strength $S(J_u,J_l)$:
\begin{equation}
    f_{J_u J_l} = \frac{1}{3(2J_l+1)} \frac{|E_{J_u} - E_{J_l}|}{h\,c\,R_M} \frac{S(J_u,J_l)}{a^2 e^2}
    \label{eq:multiplet_os}
\end{equation}

Here $a = a_0 (m_e / m_{He} + 1)$ is the scaled Bohr radius. Line strength is defined via the equation \cite{hey2017}:
\begin{equation}
    S^{1/2}(J_u, J_l) = (-1)^\phi \delta_{\kappa_u \kappa_l}\delta_{S_uS_l}[J_u,J_l,L_u,L_l]^{1/2}
    \times\begin{Bmatrix}
        L_u & J_u & S_u \\
        J_l & L_l & 1
    \end{Bmatrix}
    \begin{Bmatrix}
        l_u & L_u & L_1 \\
        L_l & l_l & 1
\end{Bmatrix} P_{l_ul_l}
    \label{eq:lin_str}
\end{equation}

Here quantum numbers in upper case represent total quantum numbers of multi electron system, while the ones in lower case represent quantum numbers of initial and final states of transiting electron. $L_1$ represent total azimuthal quantum number of all electrons without the transiting one. Phase factor is $\phi = S_u + L_1 + J_l + l_l$. In eq. (\ref{eq:lin_str}) symbol $\kappa$ represents remaining electron configuration of atom without the transiting electron. Kronecker delta symbols are here to ensure that spin and electron configuration remain unchanged during the transition. Other quantities in eq. (\ref{eq:lin_str}) are defined via the following equations:
\begin{equation}
[J_u,J_l,L_u,L_l] = (2J_u+1)(2J_l+1)(2L_u+1)(2L_l+1)    
\end{equation}

Quantity $P_{l_ul_l}$ represents transition probability:
\begin{equation}
    P_{l_ul_l} = -e\,(-1)^{l_u + l_>} \delta_{l_l, l_u\pm1} \sqrt{l_>} \,R_{n_ul_u,n_ll_l}
\end{equation}

Here $l_> = \text{max}(l_u, l_l)$. Kronecker delta ensures that transition is dipole-allowed. Value $R_{n_ul_u,n_ll_l}$ is radial integral for an electric dipole transition. In this paper we use $R_{n_ul_u,n_ll_l}$ as given in \cite{klarsfeld1989}:
\begin{multline}
    R_{n_ul_u,n_ll_l} = \frac{a_0}{Z} K_u K_l \left( \frac{\nu_u \,\nu_l}{\nu_u +\nu_l}\right)^{\nu_u + \nu_l + 2}
    \times\frac{\Gamma(\nu_u - l_l + 2)\Gamma(\nu_u + l_l + 3)}{\Gamma(\nu_u - \nu_l + 3)} \times\\
    \times\sum_{q=0}^{q_{max}} a_q\, _2F_1(-\nu_l -l_l + 1, -\nu_l -l_l, \nu_u - \nu_l + 3 - q; y)
    \label{eq:radial_int}
\end{multline}

Here $a_0$ is Bohr radius, $Z$ is electric charge, $\Gamma$ is gamma function, $_2F_1$ is Gaussian hypergeometric function. Hartree's generalized normalization factors $K_{i}$ are taken the same as in \cite{bates1949, klarsfeld1989, regem1979} and defined via the equation :
\begin{equation}
    K_i = \left( \frac{2}{\nu_i}\right)^{\nu_i} \frac{1}{\sqrt{\nu_i^2 \, \Gamma(\nu_i + l_i + 1) \, \Gamma(\nu_i - l_i)}}
\end{equation}

Hypergeometric function in eq. (\ref{eq:radial_int}) depends on parameter $y = (\nu_u - \nu_l) / (2\nu_u)$. Series coefficients $a_q$ are calculated recursively starting from $a_0=1$ using the relation taken from \cite{hey2017}:
\begin{equation}
    a_q = \frac{\nu_l - \nu_u - 3 - q}{q} \frac{\nu_u + \nu_l}{2\nu_l}
    \times\frac{l_u(l_u+1) - (\nu_u - q) (\nu_u - q + 1)}{l_l(l_l + 1) - (\nu_u - q + 2) (\nu_u - q + 3)} a_{q-1}
\end{equation}

Summation in eq. (\ref{eq:radial_int}) is truncated at $q_{max} = \text{round}(\nu_u - l_l + 1)$. Physical reasoning behind such criteria for truncation can be found in \cite{hey2017}.

We sum eq. (\ref{eq:multiplet_av}) over total angular momenta $J_u$ and $J_l$ to obtain total multiplet oscillator strength. Summation over $J$ can be easily done using equality from \cite{sobelman}:
\begin{equation}
    \sum_{J_u, J_l}(2J_u + 1)(2J_l + 1) \begin{Bmatrix}
        L_u & J_u & S_u \\
        J_l & L_l & 1
    \end{Bmatrix}^2 = 2S_u + 1
\end{equation}

Thus we obtain the following equation for oscillator strength averaged over multiplet:
\begin{equation}
    f_{ul} = \frac{1}{3} \frac{\Delta E_{ul}}{hc\,R_M} \frac{l_> (2L_u + 1)}{a^2} \begin{Bmatrix}
        l_u & L_u & L_1 \\
        L_l & l_l & 1
\end{Bmatrix}^2 R_{n_ul_u,n_ll_l}^2
\end{equation}

The remaining Wigner $6j$-symbol can be easily calculated analytically since we are modeling transitions in two-electron Helium atom. This means that the remaining electron is in the ground state, therefore total azimuthal momentum is defined solely by momentum of transiting electron: $L_u = l_u$ and $L_l = l_l$, and also $L_1 = 0$. For electric dipole transition we have $l_l = l_u \pm 1$, and using exact formulas for Wigner $6j$-symbols from \cite{varsh} we obtain the following expressions for them:
\begin{equation}
\begin{Bmatrix}
        l_u & l_u & 0 \\
        l_l & l_l & 1
\end{Bmatrix} = \frac{(-1)^{2l_u}}{2} \sqrt{\frac{(2l_u + 2)^2}{(l_u + 1)(2l_u + 1)(2l_u + 3)}}
\label{eq:wigner1}
\end{equation}

\begin{equation}
\begin{Bmatrix}
        l_u & l_u & 0 \\
        l_l & l_l & 1
\end{Bmatrix} = \frac{(-1)^{2l_u}}{2}  \sqrt{\frac{4}{(2l_u - 1)(2l_u + 1)}}  
\label{eq:wigner2}
\end{equation}

The equation (\ref{eq:wigner1}) is for the case $l_l = l_u + 1$, and the equation (\ref{eq:wigner2}) is for $l_l = l_u - 1$.

Resulting form of oscillator strength for electric dipole transitions in Coulomb approximation in Helium atom is defined via the equation:
\begin{equation}
    f_{ul} = \frac{1}{3} ~ \frac{\Delta E_{ul}}{hc\,R_M} ~\frac{l_> (2l_u + 1)}{a^2}~ W_{6j}^2~ R_{n_ul_u,n_ll_l}^2
\end{equation}
Here $a$ is again the scaled Bohr radius, $W_{6j}$ is defined via eq. (\ref{eq:wigner1}, \ref{eq:wigner2}), radial integral is defined via eq. (\ref{eq:radial_int}).

We calculate A-value from oscillator strength obtained via either of methods using eq. (6.20) from \cite{draine}:
\begin{equation}
    A_{ul} = \frac{0.6670~\text{cm$^2$ s$^{-1}$}}{\lambda^2_{ul}}\frac{g_l}{g_u}f_{ul}
\end{equation}
Here $\lambda_{ul}$ is wavelength of the  emitted photon in cm, $g_i$ is statistical weight of $i$ sublevel.

If a transition $\gamma_u \rightarrow \gamma_l$ occurs between sublevels with both $l_u>7$ and $l_l>7$ than A-values are calculated using hydrogenic approximation since sublevels with $l>7$ are essentially degenerate. The main difference from pure hydrogen transitions comes from different mass of Helium atom and its electric charge. We calculate A-values using algorithm described in \cite{sh1991}. In the paper authors present an fast and efficient iterative scheme for evaluation of hydrogenic line strength for arbitrary values of $n$ and $l$. We created a {\textit{Python 3}} program based on this scheme to calculate hydrogenic A-values.

\begin{figure}[ht]
     \centering
     \includegraphics[width=0.83\linewidth]{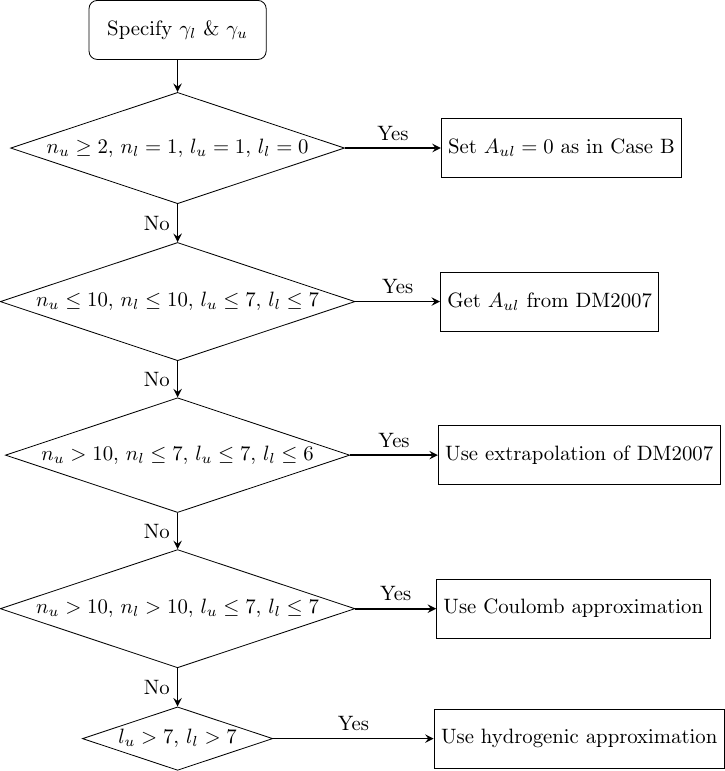}
     \caption{A flowchart describing algorithm for calculation of A-values for dipole allowed transitions in Helium atom. Abbreviation DM2007 corresponds to the paper \cite{dm2007}.}
     \label{fig:a_algorithm}
 \end{figure}

 The general algorithm for evaluation of A-values is summarized in the flowchart presented in Fig. \ref{fig:a_algorithm}.

In the calculation we use Case B approximation and thus set all A-values for transitions $n^1P \rightarrow 1^1S$ to zero. On top of A-values of the allowed transitions described above we add some important A-values for forbidden transitions between lower lying excited states. They include two-photon decay $2^1S \rightarrow 1^1S$ with $A = 50.94$ s$^{-1}$ from \cite{drake1986}, forbidden transition $2^3S \rightarrow 1^1S$ with $A = 1.27 \times 10^{-4}$ s$^{-1}$ from \cite{lp2001}, and forbidden transition $2^3P \rightarrow 1^1S$ with $A = 177.6$ s$^{-1}$ from \cite{dm2007}.

\subsection{\label{subsec:rrr}Radiative recombination rates}

Radiative recombination rates for $LS$-resolved sublevels in He\,I are calculated using eq. (14.2) from \cite{draine}:
\begin{equation}
    \alpha_\gamma(T) = \sqrt{\frac{8kT}{\pi m_e}}  \int_0^\infty \sigma_{rr, \gamma}(E)\ \frac{E}{kT}\ \exp\left(-\frac{E}{kT} \right) d\frac{E}{kT}
    \label{eq:drain_alpha}
\end{equation}

Here $T$ is temperature in K, $\sigma_{rr, \gamma}(E)$ is radiative recombination cross-section to sublevel $\gamma$.

We calculate radiative recombination rates to sublevels with $n\leq 25$ and $l\leq4$ using a direct integration of large modern set of recombination cross-sections from \cite{hs1998}. It should be noted, that in \cite{hs1998}  authors present a set of photoionization cross sections. We obtain recombination cross sections from this data using Milne relation. 

We calculate radiative recombination rates to levels with $n>25$ and/or $l>4$ via the following methods. As noted in \cite{hs1998}, there is a relation between He recombination rates and pure hydrogenic recombination rates for sublevels with $l\geq4$:
\begin{equation}
    \alpha_\gamma(T) = \frac{s}{4} \times \alpha_{\rm H}(T, n, l)
    \label{eq:hs1998}
\end{equation}
Here $s$ is multiplicity of a sublevel, and $\alpha_{\rm H}(T, n, l)$ is pure hydrogenic recombination rate calculated for sublevel with the specified values of $n,l$. We use this relation to calculate radiative recombination rates to all levels with $l\geq4$. We calculate hydrogenic recombination rates using eq. (\ref{eq:drain_alpha}), where the radiative recombination cross section is computed using algorithm described in \cite{sh1991} (similar to one used for evaluation of hydrogenic A-values).

For levels with $n>25$ and $l\leq2$ we use a generalization of eq. (\ref{eq:hs1998}):
\begin{equation}
    \alpha_\gamma(T) = f(T, \gamma) \times \alpha_{\rm H}(T, n)
\end{equation}
Here $f(T, \gamma)$ is a scaling term defined via the following equation:
\begin{equation}
    f(T, \gamma) = \frac{a_1(T)}{n^{a_2(T)}} + a_3(T)
    \label{eq:rrr_scale_eq}
\end{equation}

Here $a_i(T)$ are fit coefficients. Their dependence on temperature is determined by a fourth-degree polynomial with temperature $T$ in K:
\begin{equation}
    a_i(T) = \sum_{j=0}^4 b_j^{(i)} \left( \frac{T}{10^4~\text{K}}\right)^j
\end{equation}

\begin{figure}[ht]
    \centering
    \includegraphics[width=0.85\linewidth]{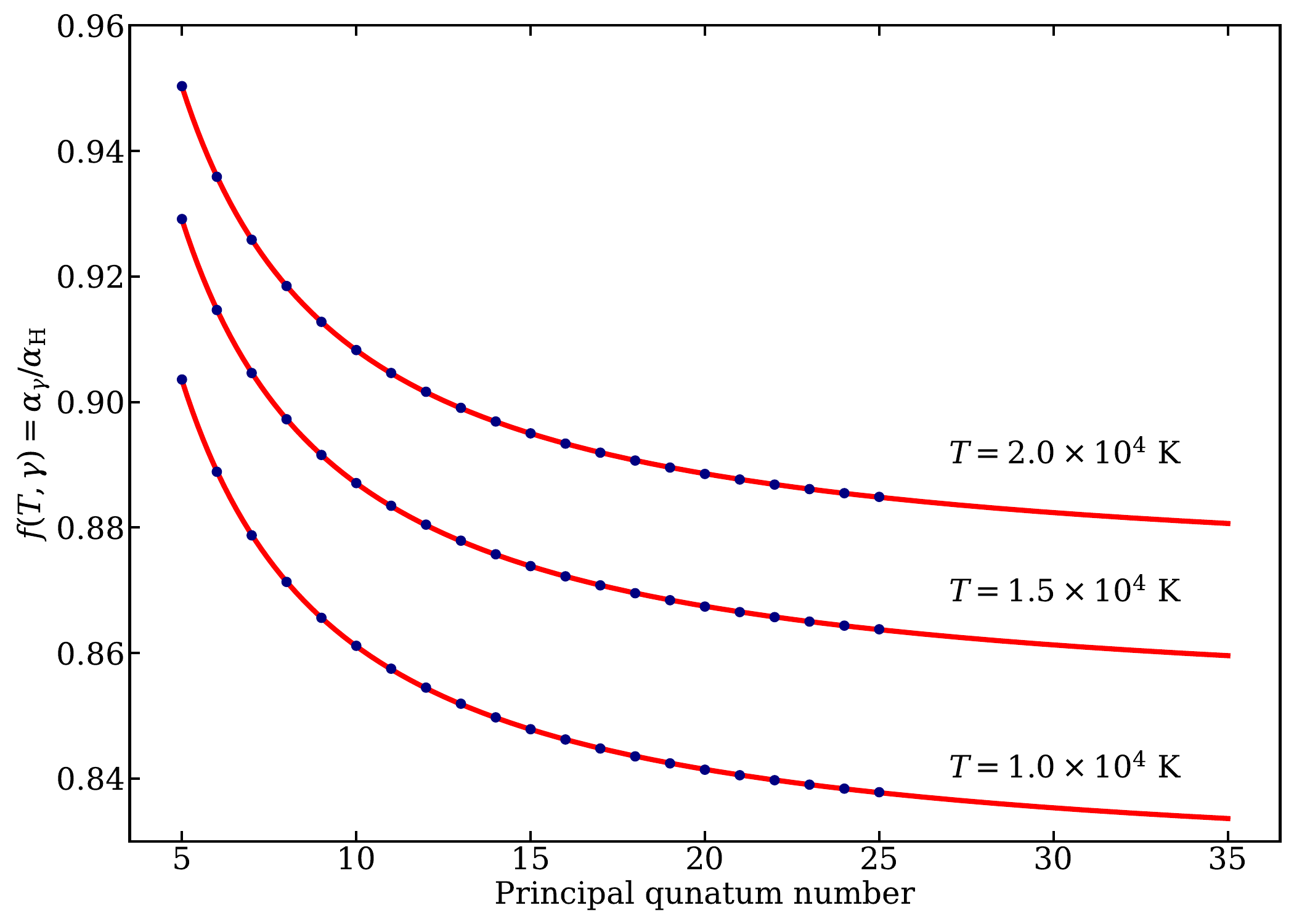}
    \caption{Scaling factor $f(T, \gamma)$ for calculation of radiative recombination rate to levels $n^1P$ for $T = 10^4$, $1.5\times10^4$ and $2\times10^4$ K. Blue circles represent exact data on ratio $\alpha_{\gamma}/\alpha_{\rm H}$. Red line represents fit defined in eq. (\ref{eq:rrr_scale_eq}).}
    \label{fig:rrr_scaling}
\end{figure}

Fit parameters $b_j^{(i)}$ are calculated using trust region reflective algorithm based on data from \cite{hs1998}. Accuracy of the fit is better than 1\% for $5000 \leq T  \leq 25000$ K. Example of such fit for radiative recombination rates to levels $n^1P$ for different temperatures is presented on Fig. \ref{fig:rrr_scaling}. Fitting coefficients for the scaling term are presented in Appendix \ref{app:rrr_coef}.

\subsection{\label{subsec:col_rates}Collisional transition rates}

Calculation of collisional rates for transitions $\gamma_i\,\rightarrow\,\gamma_j$ is one of the most challenging aspects of the model, both computationally and physically, and must be approached with great care. Several types of collisional transitions can occur, and for some of them direct and accurate atomic data are lacking. In such cases, one have to rely on approximate methods of varying accuracy. Under the physical conditions typical of the interstellar medium in H II regions, the uncertainties introduced by these approximations are expected to be small, since the electron densities in such environments are relatively low ($\sim 10\text{--}1000$ cm$^{-3}$). Nevertheless, this does not imply that collisional transitions can be neglected, as they have a substantial impact on many important optical and near-infrared He\,I lines (e.g., $\lambda7065$ and $\lambda10830$). The influence of including collisional transitions on the resulting He\,I line emissivities can be assessed by comparing the models of He\,I spectrum from \cite{bauman2005} and \cite{porter2005}. The former is a pure radiative--recombination model in which collisions are not included, whereas the latter is a full collisional--recombination model. In the collisionless model of \cite{bauman2005}, the resulting emissivities are generally lower by $1\text{--}2\%$ compared to those obtained in \cite{porter2005}. Lines strongly affected by collisional transitions exhibit even larger differences; for example, the emissivity of $\lambda10830$ is $18\%$ lower in the collisionless model. This clearly demonstrates the importance of rigorous treatment of collisional processes in helium atomic models.

From a computational standpoint, implementing a full collisional scheme is challenging, since the number of possible collisional transitions between sublevels in the helium atom increases rapidly with the number of modeled levels: each sublevel is collisionally coupled to all other sublevels.

In our He\,I emission model we distinguish several types of collisional transitions. Each type of transitions is treated by a dedicated method, as described below.

For transitions $\gamma_l \rightarrow \gamma_u$ between sublevels with $n\leq5$ we calculate collisional excitation (transition $l\rightarrow u$) and de-excitation (transition $u\rightarrow l$) rates using the standard formulae from \cite{osterbrock}:

\begin{equation}
    q_{lu} = \frac{8.629\times10^{-6}}{\sqrt{T}} \frac{\Upsilon_{ul}(T)}{g_l} \, \exp \left ( -\frac{\Delta E_{ul}}{kT}\right)
\end{equation}

\begin{equation}
    q_{ul} = \frac{8.629\times10^{-6}}{\sqrt{T}} \frac{\Upsilon_{ul}(T)}{g_u} 
\end{equation}

Here $T$ is temperature in K, $g_l$ and $g_u$ are statistical weights of the corresponding sublevels, $\Upsilon_{ul}(T)$ is effective collision strength of the transition. We calculate effective collision strength using direct integration of collision strength $\Omega(E)$ for the specified temperature $T$:
\begin{equation}
    \Upsilon_{ul}(T) = \int_0^\infty \Omega(E) \exp\left( -\frac{E}{kT}\right) d\left(\frac{E}{kT}\right)
\end{equation}

Integration is carried out over an energy grid using the linear interpolation method described in \cite{burgess1992}. The tabulated values of collision strengths as a function of energy $E$ were derived from high-resolution collisional cross-section data. The cross-section data was computed using the convergent close-coupling (CCC) method \cite{bray1994, fursa1995} and provided by Igor Bray (private communication). This constitutes an important distinction between our helium model and the previous studies. Other independent studies \cite{porter2005, porter2012, delzanna2020, delzanna2022} rely on collisional data from \cite{bray2000}, which is a widely used source of atomic data for collisional transitions between sublevels with $n_l \leq 2$ and $n_u \leq 5$ in the helium atom. In the paper, the authors present extensive tables of thermally averaged collision strengths on a temperature grid. In previous helium atom models, the effective collision strengths are obtained via linear interpolation on this grid, which is relatively sparse in very important range $\sim 10^4 - 2\times 10^4$ K. Furthermore, for transitions not included in \cite{bray2000}, other studies adopt values from \cite{sawey1993}, where a different computational method was used and the data are presented on a different temperature grid. The complementing dataset from \cite{sawey1993} covers transitions between levels up to $n = 4$.

Rather than combining heterogeneous datasets and using the linear interpolation approach applied in earlier works, we integrate the collision strengths directly. This yields a more accurate and internally consistent treatment of collisional transitions between sublevels up to $n = 5$ over the full range of temperatures typical for the interstellar medium of H II regions.

For transitions between sublevels $n_l\leq 5$ and $n_u > 4$ we use scaled collision strengths calculated via the relation from \cite{ralchenko2008}:
\begin{equation}
    \Omega(\gamma_l, \gamma_u, E) = \frac{f_{\gamma_l, \gamma_u}}{f_{\gamma_l, \gamma_5}} \,\Omega(\gamma_l, \gamma_5, E)
    \label{eq:scaled_cs}
\end{equation}

Here $\gamma_5 = [5, l_u, s_u]$. Oscillator strengths are calculated as described in Sec. \ref{subsec:a_vals}. In case there is no oscillator strength available for a transition (this is the case for e.g. dipole-forbidden transitions), we replace ratio $f_{\gamma_l, \gamma_u}/f_{\gamma_l, \gamma_5}$ with ratio $(5 / n_u)^3$ which follows from Born approximation. 

For transitions between sublevels with $n_l\geq5$ and $n_u\,\geq\,5$  there are two possible types of transitions which are briefly described below. It is important to note that both of these processes are extremely important because they redistribute the populations of Rydberg states toward LTE. 

Transitions occurring between sublevels with the same principal quantum number are called $l$-changing or Stark collisions. Such collisions modify the orbital angular momentum $l$ of an electron within a fixed principal quantum number $n$ through Stark mixing induced by the electric fields of protons or other heavy charged particles \cite{ps1964, vos2001}. The interaction with the electric field of a heavy particle lifts the energy degeneracy of highly excited sublevels within the $n$-shell, enabling collisional transitions to take place. Since the energy separation between these sublevels is small, even a low-energy heavy particle can induce a change from one sublevel to another. The slower the particle -- the longer the interaction time, and therefore at lower temperatures the probability of transitions induced by heavy particles increases due to the prolonged action of the particle’s electric field on the atomic shell. For $l$-mixing collisions main colliders are heavy particles like protons and ions of He. A great in-depth review of the underlying physics of the process and ways to evaluate the corresponding rates can be found in \cite{guzman1, guzman2, guzman4}.

Transitions between sublevels with different principal quantum numbers are referred to as $n$-changing. Since levels with different $n$ are separated by significant energy gaps, only fast-moving electrons can induce collisional transitions between them. A comprehensive review of the underlying physics, along with a comparison of different approximations for collisional rates, is provided in \cite{guzman3}.

We compute the rates of $l$-changing collisions using the \textit{Lmixing} package for \textit{Python 3} \cite{lmixing}. This package is designed to calculate the rate coefficients for angular momentum mixing of Rydberg atoms in collisions with heavy particles in astrophysical plasmas. In the model we consider two types of heavy particles: protons and single-ionized helium atoms with number density $n_{He^+} = 0.1\,n_p$. Following the recommendations from the authors of the package, we employ the full quantum-mechanical transition probabilities from \cite{vos2001, vos2012} to evaluate dipole transition rates for levels with $n \leq 30$. For levels with $n > 30$, we adopt the semiclassical approach of \cite{vos2017}. A limitation of the semiclassical method, however, is that it operates within an unresolved framework and returns only the total collision rate:
\begin{equation}
    q_{nl} = q_{nl, nl+1} + q_{nl, nl-1} \equiv q_{nl}^+ + q_{nl}^-
\end{equation}

We obtain resolved rates from the unresolved ones using the reciprocity condition from \cite{guzman4}:
\begin{equation}
    q_{nl}^+ = \frac{2l + 3}{2l + 1}q_{nl}^-
\end{equation}

Rates are obtained iteratively starting from the sublevel with $l=n-1$, since there is no sublevel with $l=n$ and thus $q_{nn-1}^+ = 0$ and $q_{nn-1} = q_{nn-1}^-$.

In addition to the dipole $l$-changing collisions we incorporate quadrupole $l$-changing collisions in our model. Rate of these collisions is calculated using the eq. (C1) from \cite{Deliporanidou}.

We calculate rate of $n$-changing collisions using straight trajectory Born approximation using method from \cite{lebedev1998} as suggested in \cite{guzman3}. The rate is calculated via the equation:
\begin{equation}
    q_{n'n} = \frac{g_n}{g_{n'}} 2\sqrt{\pi} a_0^2 \alpha \,c\,n \left[ \frac{n'}{Z(n'-n)}\right]^3 \frac{f(\theta) \phi}{\sqrt{\theta}}
\end{equation}

Here $g_n$ and $g_{n'}$ are statistical weights of levels, $a_0$ is Bohr radius, $\alpha$ is fine structure constant. Variable $\theta = kT / (Z^2 I_H)$, and $f(\theta)$ is given by:
\begin{equation}
    f(\theta) = \frac{\ln \left( 1 + \frac{n\theta}{Z(n' - n)\sqrt{\theta} + 2.5}\right)}{\ln \left(  1 + \frac{n\sqrt{\theta}}{Z(n' - n)} \right)}
    \label{eq:nchange}
\end{equation}

Function $\phi$ is defined via the following equation:
\begin{eqnarray}
    \phi &&= \frac{2 n'^2 n^2}{(n' + n)^4 (n'-n)^2}
    \times\left[ 4(n' - n) - 1\right]\exp\left( \frac{E_n}{kT}\right) E_1\left( \frac{E_n}{kT}\right) 
    + \frac{8n^3}{(n' + n)^2 (n' - n) n^2 n'^2}\nonumber\\
     &&\times(n' -n -0.6) \left( \frac{4}{3} + n^2(n'-n)\right) 
    \times \left[ 1 - \left( \frac{E_n}{kT}\right) \exp\left( \frac{E_n}{kT}\right) E_1\left( \frac{E_n}{kT}\right)\right]
\end{eqnarray}

Here $E_n$ is energy of $n$-level, $E_1$ is the first exponential integral. The drawback of rate defined via eq. (\ref{eq:nchange}) is that it returns unresolved collision rate. To obtain resolved rates we average over the statistical weight of the initial and final levels as in \cite{guzman5}:
\begin{equation}
    q_{nl, n'l'} = \frac{2l' + 1}{n'^2} q_{nn'}
\end{equation}

\begin{figure}[ht]
    \centering
    \vspace{0.5cm}
    \includegraphics[width=0.83\linewidth]{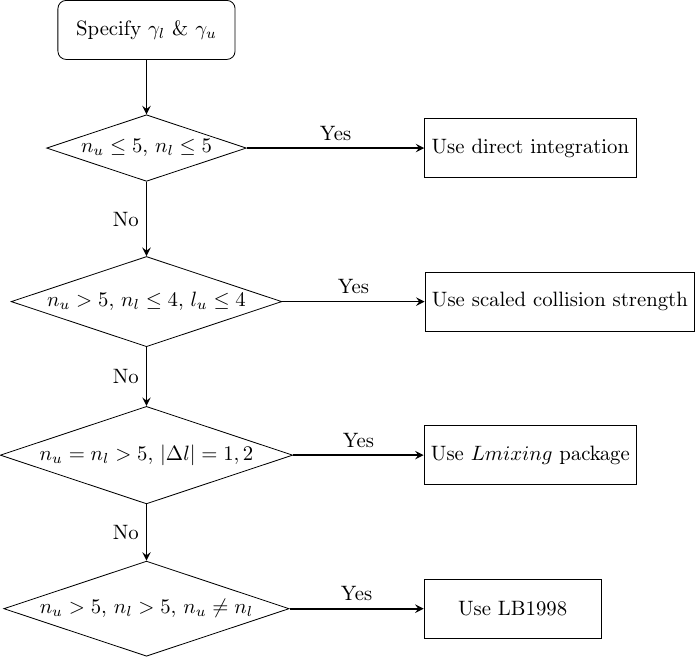}
    \caption{A flowchart describing algorithm for calculation of collisional rates in Helium atom. Abbreviation LB1998 corresponds to collision rate from \cite{lebedev1998}.}
    \label{fig:q_algorithm}
\end{figure}

 The general algorithm for calculation of collision rates is summarized in the flowchart presented in Fig. \ref{fig:q_algorithm}.

\subsection{\label{subsec:col_ion}Collisional ionization rates}

Collisional ionization rates are calculated using equation:
\begin{equation}
    C^{(ion)}_{\gamma}(T) = \sqrt{\frac{8kT}{\pi m_e}} \int_{I_{\gamma}}^\infty \sigma^{(ion)}_{\gamma}(E)\frac{E}{kT}\exp\left(-\frac{E}{kT} \right) d\frac{E}{kT}
\end{equation}

Cross sections for collisional ionization were taken from \cite{ralchenko2008}. In the paper authors provide analytical formulas for the collisional ionization cross sections for levels with $n\leq4$ which are easily integrable. For levels with $n>4$ we use the following extrapolation from \cite{ralchenko2008}:
\begin{equation}
    \sigma^{(ion)}_{n>4} = 2.32 \times 10^{-16} \left( \frac{Ry}{E_{n>4}}\right)^2 \,\left( \frac{x-1}{x^2} \right) \, \ln(1.25x) ~~\text{cm}^2
\end{equation}
Here $Ry$ is the Rydberg energy, variable $x = E / E_{n>4}$, and $E$ is kinetic energy of the colliding electron.

In the model we do not account for dielectronic recombination since for the typical physical conditions of HII regions the rate of this process is by 3-4 orders of magnitude lower compared to radiative recombination rate as shown in \cite{delzanna2020}.

\subsection{\label{subsec:esc_prob}Optical depth and photon escape probability}

The transition $2^3S \rightarrow 1^1S$ is strongly forbidden with $A = 1.27 \times 10^{-4}$ s$^{-1}$ \cite{lp2001}. The $2^3S$ sublevel is metastable, and an absorption of photons from $n^3P \rightarrow 2^3S$ transitions may become quite a significant effect when modeling emissivities of Helium lines. This is due to the long lifetime of the $2^3S$ sublevel which leads to the emergence of a non-zero optical depth in such lines. We account for radiative transfer effects by replacing A-values for transitions $n^3P \rightarrow 2^3S$ with the expression:
\begin{equation}
    A(n^3P \rightarrow 2^3S) \longrightarrow \epsilon(\tau) A(n^3P \rightarrow 2^3S)
\end{equation}
Here $\epsilon(\tau)$ is mean photon escape probability, $\tau$ is optical depth in the specified line. We calculate escape probability using the formula from \cite{bss2002}:
\begin{equation}
    \epsilon(\tau) = \frac{1.72}{1.72 + \tau}
\end{equation}

Since optical depth in every transition $n^3P \rightarrow 2^3S$ is defined by population of metastable $2^3S$ sublevel, it is convenient to parametrize optical depth of any line in terms of optical depth of some reference line. As a reference line we use $\lambda3889$ line which corresponds to the transition $3^3P \rightarrow 3^3S$ (same as done in \cite{bss2002}). Optical depth $\tau_{3889} \equiv \tau$ is an additional parameter in the model of Helium atom alongside electron density $n_e$ and temperature $T$. Optical depths of other lines and the corresponding escape probabilities are calculated using the relation:
\begin{equation}
    \tau_\lambda = \frac{\lambda^2}{3889^2} \frac{A_\lambda}{A_{3889}} \tau
\end{equation}

\subsection{\label{subsec:rec_rem}Recombination reminder}

If one aims to compute the exact population of a given sublevel in the helium atom (or any other atom), then, strictly speaking, an infinite number of levels would need to be modeled. This is, of course, impossible in practice. Instead, the model must be truncated at some maximum principal quantum number $n_{\max}$, while the contribution of levels with $n > n_{\max}$ must be accounted for separately. Radiative and collisional transitions from such highly excited levels provide an additional and non-negligible channel for populating lower states.  

We account for this effect following the approach of \cite{bauman2005}. Specifically, we introduce the recombination reminder $\alpha_{\mathrm{rem}}$ as an explicit addition to the direct recombination of the $LS$-resolved sublevels with $n = n_{\max}$. The recombination reminder is defined as the convergent sum of radiative recombination rates into levels with $n > n_{\max}$. We evaluate $\alpha_{\mathrm{rem}}$ using the method of \cite{seaton1959}:
\begin{equation}
    \alpha_{rem}(n_{\max}, T) = \sum_{n = n_{\max}}^{\infty} \alpha_n = \frac{1}{2} D Z \sqrt{\lambda } \left[ \frac{x_{n_{\max}}}{n_{\max}} S_{n_{\max}}(\lambda) + \sigma_{n_{\max}}(\lambda) \right]
\end{equation}

Here $n_{\max}$ is the maximum modeled level, $Z$ is the electric charge, and other quantities are defined via the following equations:
\begin{equation}
\begin{aligned}
    D &= \frac{64}{3} \sqrt{\frac{\pi}{3}} \alpha^4 c a_0^2 = 5.197 \times 10^{-14} ~\text{cm$^3$ s$^{-1}$} \\
    \lambda& = \frac{hc\, \text{Ry}\, Z^2}{kT} = 157890 \times \frac{Z^2}{T} ~~\text{where $T$ in K} \\
    &~~~~~~~~~~~~~~~~~~~ x_n = \frac{\lambda}{n^2}
\end{aligned}
\end{equation}

Functions $S_n$ and $\sigma_n$ are determined via the following equations:
\begin{equation}
    S_n(\lambda) = \int_0^\infty \frac{g_{II}(n, \epsilon) \exp(-x_n u)}{1 + u} du~\text{,~ $u = n^2\epsilon$}
\end{equation}

\begin{equation}
    \sigma_n(\lambda) = \int_0^{x_n} S_{n'}(\lambda) dx_{n'}
\end{equation}

Recombination reminder calculated in this way is then explicitly added to the direct recombination rates of sublevel in $n_{\max}$ level according to their statistical weights:
\begin{equation}
    \alpha(n_{\max}, l, s, T) \longrightarrow
    \alpha(n_{\max}, l, s, T) + \frac{(2l + 1)(2s+1)}{4 \,n^2_{max}} \alpha_{rem}(n_{\max}, T)
    \nonumber
\end{equation}

\section{\label{sec:he_model}Model of He I emission spectrum}

We calculate populations of sublevels in Helium atom as a function of the following parameters: $n_e$ - electron density, $T_e$ - temperature, and $\tau$ - optical depth in $\lambda3889$ line. General form of detailed balance equation for population of sublevel $\gamma_k$ follows the standard form, that could be found in e.g. \cite{bss1999}:
\begin{eqnarray}
    n_en_{\rm He} \alpha_k + \sum_{i>k} n_i A_{ik}(\tau) 
    + \sum_X \sum_{j\neq k} n_X n_j q_{jk} = \nonumber\\
    n_k \left[n_e C^{(ion)}_k + \sum_{i<k} A_{ki}(\tau)
     + \sum_X \sum_{j\neq k} n_X q_{kj}\right]
\end{eqnarray}

In this equation left-hand part includes all processes that populate sublevel $\gamma_k$, and right-hand part includes all depopulation processes. Population processes include radiative recombination, radiative decay from upper levels relative to $\gamma_k$, and collisional transitions of all types from all other levels. Depopulation processes include collisional ionization, radiative decays to lower levels, and again collisional transitions to all other levels from level $\gamma_k$. Details on calculation of atomic data are provided in Sec. \ref{sec:at_data}. We write down similar equations for every sublevel under consideration and solve the obtained system for sublevel populations using matrix inversion method.

A cornerstone in constructing such systems of detailed balance equations is the choice of the maximum number of explicitly modeled sublevels. Previous studies adopted different truncation limits, ranging from $n_{\max}=5$ in \cite{bss1999} to as many as 40 $LS$-resolved and 100 $n$-resolved levels on top of the $LS$-resolved manifold in \cite{delzanna2022}. In this paper, we chose number of modeled levels according to the methodology proposed in \cite{guzman5}. The authors provide simple criteria to estimate the maximum number of $LS$-resolved Rydberg states for which radiative decay rates still dominate over collisional ones. Once collisional processes become faster than radiative decays, the corresponding levels are assumed to be populated according to their statistical weights.  

In addition, \cite{guzman5} investigated the number of explicitly modeled $LS$-resolved levels required to achieve convergence of the full radiative–recombination cascade. Their results show that, within the relevant range of densities and temperatures, convergence at the $\leq 1\%$ level requires $n_{\max} \gtrsim 25$. To guide the choice of $n_{\max}$, the authors propose the following rule of thumb:
\begin{equation}
    n_{\max} = n^* + 10,
\end{equation}
where $n^*$ is the principal quantum number of the level for which the critical electron density $n_{crit}$ equals the model parameter $n_e$. The critical density is estimated using the empirical relation from \cite{guzman5}:
\begin{equation}
    n_{\mathrm{crit}} = 7.599 \times 10^{13}\, \cdot \, n^{-8.675}
\end{equation}

We tested the convergence of our model as a function of the number of explicitly modeled sublevels and found good agreement with the results of \cite{guzman5}. To ensure robustness and avoid potential convergence issues, we adopted a conservative choice of $n_{\max} = 50$, which exceeds the value recommended by the empirical criterion from \cite{guzman5}.

The populations of $LS$-resolved levels and the emissivities of spectral lines at given values of temperature $T$, electron density $n_e$, and optical depth $\tau$ are calculated using a dedicated \textit{Python 3} code, which we developed specially for this problem. In order to improve performance efficiency of the model we have pre-computed all needed A-values and oscillator strengths for all transitions under consideration and stored them in separate files. Additionally, we have done the same for all energies and effective quantum numbers of $LS$-resolved sublevels up to specified $n_{\max}$. It allowed to save about 70\% of computational time per a combination of $n_e$, $T$ and $\tau$ parameters. The total execution time for calculating the sublevel populations was $\sim$ 150-170 seconds for one set of parameters on AMD Ryzen 7 8845HS CPU in single thread mode. Testing showed that majority of computational time is spent on calculation of $n$-changing collisional rates.

\section{\label{sec:results}Results}

\subsection{\label{subsec:emissivities}Emissivities of He I lines}

We compute the populations of all $LS$-resolved sublevels of the helium atom up to $n_{\text{max}} = 50$ using the model described above, for the following grids of electron densities and temperatures. For electron densities, we adopt a linear grid in the range $1 \leq n_e \leq 500$ cm$^{-3}$ with increments of 10 cm$^{-3}$, and a logarithmic grid with steps of 0.05 dex in the range $500 \leq n_e \leq 1000$ cm$^{-3}$. For range $1000 \leq n_e \leq 10000$ cm$^{-3}$ we adopt logarithmic grid with increments of 0.25 dex. The fine spacing in the low-density range is motivated by observational estimates of electron densities in compact blue dwarf galaxies, which typically fall within $\sim 10 - 300$ cm$^{-3}$. Modern photoionization models used to infer helium abundances in H II regions \cite{he_2, he_5, he_6, he_8, he_9, he_10} employ bilinear interpolation over emissivity grids; hence, a finer density grid improves the accuracy and reliability of the derived observational constraints. For the temperature, we adopt linear grid spanning $8000 \leq T \leq 22000$ K with increments of 250 K.

The emissivity of a helium line $\lambda$ corresponding to a transition from sublevel $\gamma_i$ to sublevel $\gamma_j$ is calculated as
\begin{equation}
\frac{4\pi j_\lambda}{n_e n_{\text{He}^+}} (n_e, T) = \frac{n_i(n_e, T)}{n_e n_{\text{He}^+}} A_{ij} E_{ij},
\end{equation}

Here $n_i(n_e, T)$ is the population of the upper sublevel $\gamma_i$, $A_{ij}$ is the Einstein coefficient for the transition, and $E_{ij}$ is the energy difference between the initial and final sublevels. Thus, the density and temperature dependence of line emissivities is governed by the corresponding dependence of the sublevel populations, which are obtained from the helium atom model at a given values of $n_e$ and $T$.

We compare our resulting emissivities with the ones presented in literature for a set of so-called ``benchmark'' lines. This set includes the strongest lines ranging from near-UV to NIR ranges, and is often used in literature to compare different models of Helium recombination spectrum. The results for $n_e = 100$ cm$^{-3}$, $T=10000$ K and $n_e = 100$ cm$^{-3}$, $T=20000$ K are presented in Tables \ref{tab:emis_1e4} and \ref{tab:emis_2e4}. These tables are directly comparable to the Tables 1 and A3 from \cite{delzanna2022}.

\begin{table}[ht]
\centering
\caption{\label{tab:emis_1e4}
Emissivities ($10^{-26}$ erg cm$^3$ sec$^{-1}$) of the strongest Helium lines, for $n_e = 10^2$ cm$^{-3}$ and $T=10000$ K. Abbreviations: BSS99 stands for \cite{bss1999}, P12 for \cite{porter2012}, D-Z22 for \cite{delzanna2022}.}
\begin{tabular}{cccccc}

    \hline
    $\lambda$ & Terms & BSS99 & P12 & D-Z22 & \textbf{This work}\\
    \hline
    2945 & $5^3P - 2^3S$ & 2.70 & 2.70 & 2.70 & 2.70\\
    3188 & $4^3P - 2^3S$ & 5.62 & 5.62 & 5.61 & 5.62\\
    3889 & $3^3P - 2^3S$ & 13.7 & 14.0 & 14.0 &  14.0\\
    3965 & $4^1P - 2^1S$ & 1.39 & 1.42 & 1.41 & 1.41 \\
    4026 & $5^3D - 2^3P$ & 2.86 & 2.91 & 2.91 & 2.92\\
    4388 & $5^1D - 2^1P$ & 0.76 & 0.77 & 0.77 & 0.77\\
    4471 & $4^3D - 2^3P$ & 6.16 & 6.12 & 6.13 & 6.14\\
    4713 & $4^3S - 2^3P$ & 0.65 &0.64 & 0.65 & 0.65\\
    4922 & $4^1D - 2^1P$ & 1.64 & 1.65 & 1.65 & 1.66\\
    5016 & $3^1P - 2^1S$ & 3.49 & 3.56 & 3.55 & 3.55\\
    5876 & $3^3D - 2^3P$ & 16.9 & 16.9 & 17.0 & 16.9\\
    6678 & $3^1D - 2^1P$ & 4.79 & 4.80 & 4.83 & 4.79\\
    7065 & $3^3S - 2^3P$ & 2.96 & 2.98 & 2.98 & 2.97\\
    7281 & $3^1S - 2^1P$ & 0.90 & 0.90 & 0.90 & 0.90\\
    10830 & $2^3P - 2^3S$ & 34.0 & 33.4 & 33.6 & 33.6\\
    18685 & $4^3F - 3^3D$ & 2.22 & 2.20 & 2.23 & 2.18 \\
    20587 & $2^1P - 2^1S$ & 4.13 & 4.15 & 4.17 & 4.16\\
    \hline
\end{tabular}
\end{table}

\begin{table}[ht]
\caption{\label{tab:emis_2e4}
Emissivities ($10^{-26}$ erg cm$^3$ sec$^{-1}$) of the strongest Helium lines, for $n_e = 10^2$ cm$^{-3}$ and $T=20000$ K. Abbreviations: BSS99 stands for \cite{bss1999}, P12 for \cite{porter2012}, D-Z22 for \cite{delzanna2022}.}
\centering
    \begin{tabular}{cccccc}
    \hline
    $\lambda$ & Terms & BSS99 & P12 & D-Z22 & \textbf{This work}\\
    \hline
    2945 & $5^3P - 2^3S$ & 1.68 & 1.69 & 1.68 & 1.69\\
    3188 & $4^3P - 2^3S$ & 3.53 & 3.50 & 3.48 & 3.50\\
    3889 & $3^3P - 2^3S$ & 8.53 & 8.61 & 8.59 & 8.62\\
    3965 & $4^1P - 2^1S$ & 0.81 & 0.84 & 0.82 & 0.83\\
    4026 & $5^3D - 2^3P$ & 1.44 & 1.48 & 1.47 & 1.49\\
    4388 & $5^1D - 2^1P$ & 0.38 & 0.39 & 0.38 & 0.38\\
    4471 & $4^3D - 2^3P$ & 3.11 & 3.05 & 3.03 & 3.05\\
    4713 & $4^3S - 2^3P$ & 0.51 & 0.49 & 0.49 & 0.49\\
    4922 & $4^1D - 2^1P$ & 0.80 & 0.82 & 0.80 & 0.80\\
    5016 & $3^1P - 2^1S$ & 2.01 & 2.06 & 2.03 & 2.04\\
    5876 & $3^3D - 2^3P$ & 8.06 & 8.13 & 8.00 & 7.98\\
    6678 & $3^1D - 2^1P$ & 2.18 & 2.30 & 2.18 & 2.18\\
    7065 & $3^3S - 2^3P$ & 2.22 & 2.18 & 2.18 & 2.18\\
    7281 & $3^1S - 2^1P$ & 0.61 & 0.61 & 0.61 & 0.61\\
    10830 & $2^3P - 2^3S$ & 24.6 & 23.8 & 23.6 & 24.0\\
    18685 & $4^3F - 3^3D$ & 0.92 & 0.94 & 0.92 &  0.90\\
    20587 & $2^1P - 2^1S$ & 2.24 & 2.29 & 2.24 & 2.25\\
    \hline
    \end{tabular}
\end{table}

The comparison of emissivities presented in the Tables shows excellent agreement between our results and those of previous studies. The average difference with the emissivities reported in \cite{porter2012} for the majority of lines is $\sim 0.32\%$ at $T=10000$ K and $\sim 1.16\%$ at $T=20000$ K, while the corresponding differences with \cite{delzanna2022} are $\sim 0.32\%$ and $\sim 0.57\%$, respectively. The observed discrepancies arise from the use of different atomic datasets for the calculation of collisional rates, as already noted in Sec. \ref{subsec:col_rates}. 

\begin{figure}[htbp]
    \centering
    \includegraphics[width=0.9\linewidth]{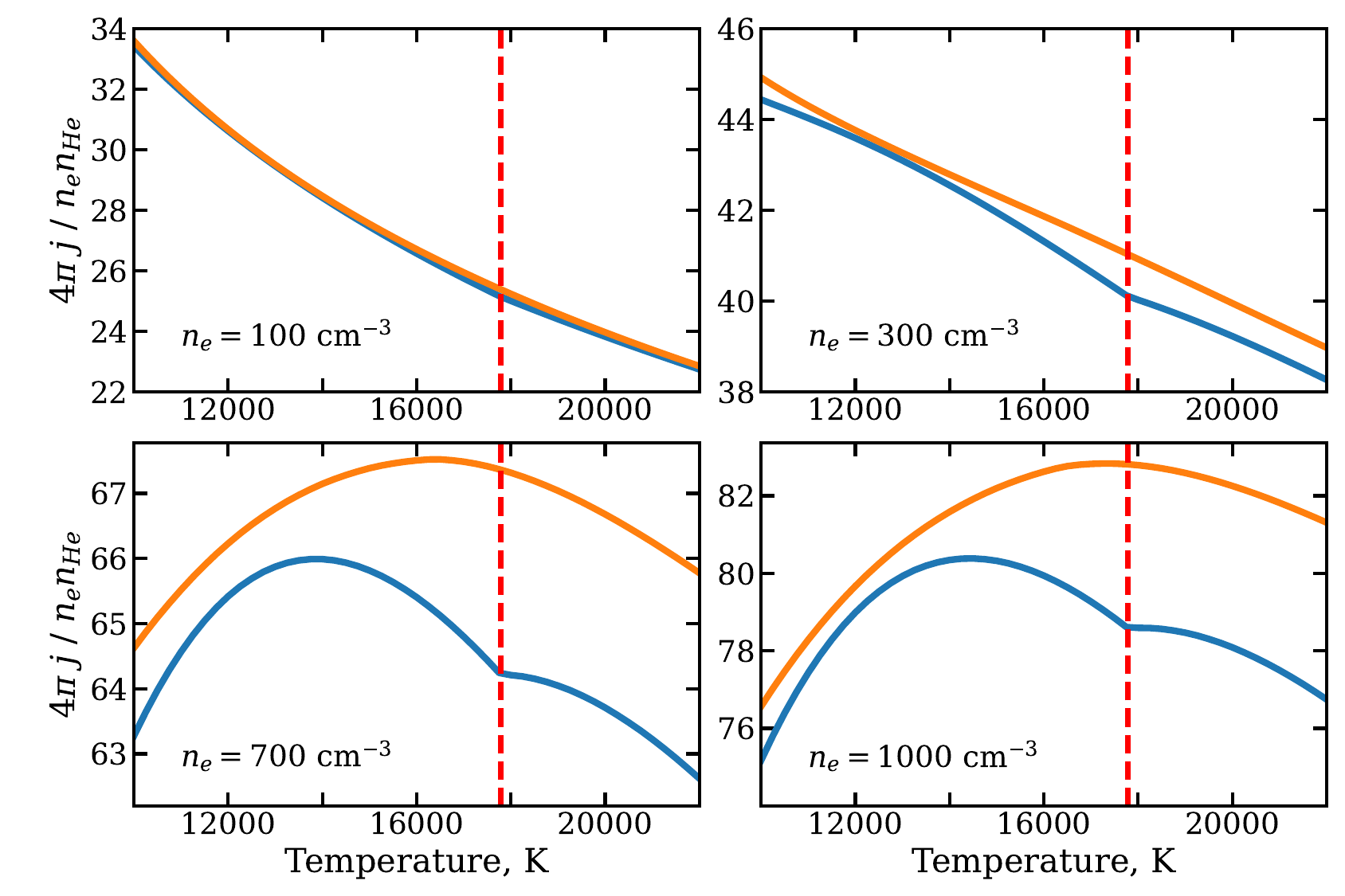}
    \caption{Emissivities of He\,I $\lambda10830$ line in units $10^{-26}$ erg cm$^3$ sec$^{-1}$, calculated for different electron densities as a function of temperature. Orange line represents emissivity calculated using our model. Blue line represents emissivity from \cite{porter2012, aver2013}. Red dashed line represents $T = 10^{4.25} = 17783$ K corresponding to interpolation node on the grid of averaged collision strengths from \cite{bray2000} (see text).}
    \label{fig:porter_10830}
\end{figure}

It is important to note that a significant systematic discrepancy was found when comparing the emissivities of the He\,I $\lambda10830$ line with the results of \cite{porter2012, aver2013}. The emissivity of $\lambda10830$ from these papers is systematically underestimated, with the magnitude of the offset growing as the electron density $n_e$ increases. In addition, the tabulated emissivities presented in \cite{aver2013} (computed using the model from \cite{porter2012} on a much finer parameter grid) exhibit a visible kink, which becomes more pronounced at higher $n_e$. These effects are illustrated in Fig.~\ref{fig:porter_10830}. The position of the kink nearly coincides with one of the grid nodes used for linear interpolation of averaged collision strengths from \cite{bray2000}: the grid node corresponds to $T = 10^{4.25} = 17783$ K, while the kink appears at $T = 17750$ K. At lower densities ($n_e \lesssim 100$ cm$^{-3}$) discrepancy between results is not very noticeable. However, as the density increases - so does the discrepancy. This suggests that in the model of \cite{porter2012} there is a possible issue with the algorithm used to compute collisional rates for transitions involving $2^3P$ level, which is the upper level of the transition corresponding to the $\lambda10830$ line. Similar problem with $\lambda10830$ emissivity from \cite{porter2012,aver2013} was previously noted and discussed in \cite{delzanna2022}, where the authors concluded that the differences stem from variations in the treatment of atomic data for collisional rates between low-lying excited levels. Since the $\lambda10830$ line is crucial in modern determinations of the primordial $^4$He abundance -- being a key diagnostic of the electron density in galaxies -- it must be recognized that the use of emissivity of $\lambda 10830$ line from \cite{porter2012, aver2013} may introduce a systematic bias into the results.

More pronounced deviations are found when comparing our results with those of \cite{bss1999}, where the differences reach $\sim 1$–$5\%$ for some lines. This highlights the impact of updated atomic data underlying our model, since all relevant atomic processes included here and in \cite{bss1999} have been recalculated with newer datasets. This distinction is particularly important, as the present research is focused on radiative transfer corrections for helium lines, while the previously adopted correction function was based on the modified model of \cite{bss1999}. Accordingly, we may expect similar differences to appear in the optical depth correction function $f_\tau$, that would be derived in next subsection.

The complete set of helium emissivities computed over the specified $n_e$ and $T$ grids is too extensive to be published in full, but is available from Zenodo \cite{zenodo}.

\subsection{\label{subsec:odf}Optical depth correction function}

We calculate optical depth correction function using the equation:
\begin{equation}
    f_\tau(\lambda, n_e, T) = \frac{E_\lambda(\tau\neq0, n_e, T)}{E_\lambda(\tau=0, n_e, T)}.
\end{equation}

Correction function was calculated on the same grid of temperatures and densities as the emissivities. For optical depth grid we chose $0 \leq \tau \leq 10$ with 0.5 increments.

\begin{figure}[htbp]
    \centering
    \includegraphics[width=0.95\linewidth]{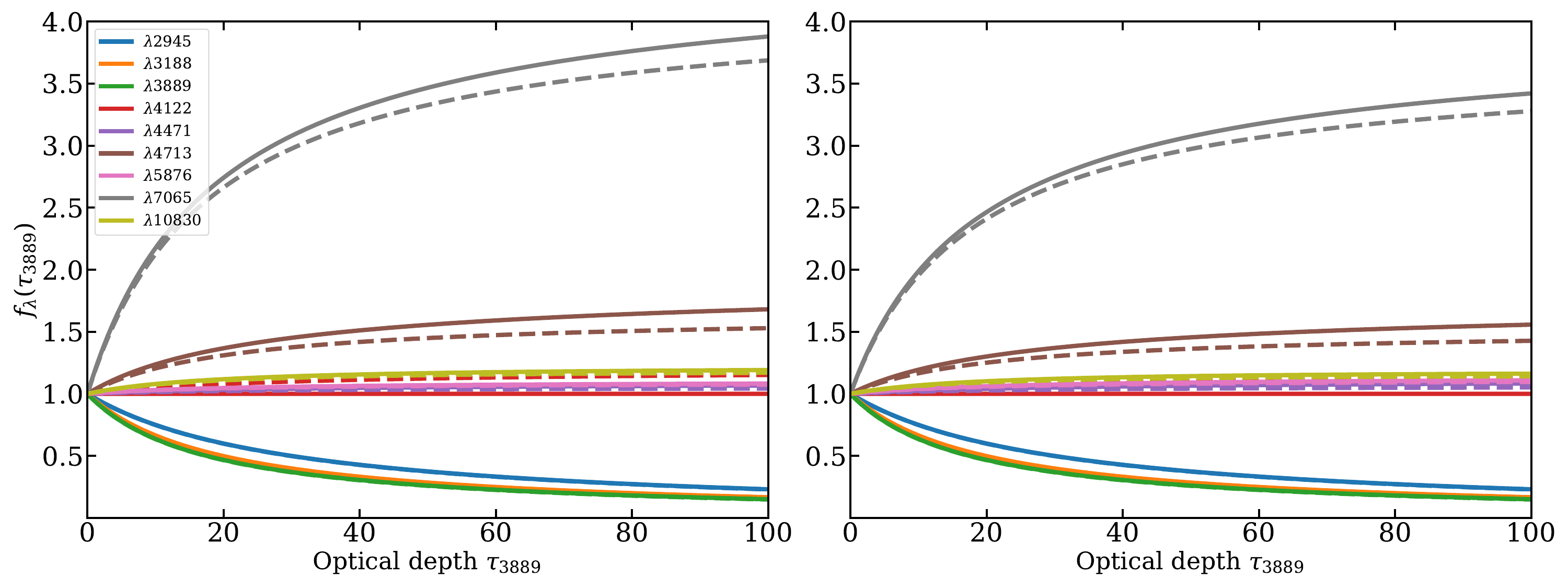}
    \caption{Optical depth correction function for selected optical lines for cases with electron density $n_e=100$ cm$^{-3}$ and temperature $T = 10000$ K (left panel) and 20000 K (right panel). Solid lines represent correction function calculated using our model of He\,I recombination spectrum, while dashed lines represent results of \cite{bss2002}.}
    \label{fig:odf1}
\end{figure}

We compare our results with the results presented in \cite{bss2002}. To do that, we fully recreated the model of He atom described in papers \cite{bss1999, bss2002} and repeated the entire cycle of calculations from the papers. We compare our results with those of \cite{bss2002} in two ways. The first is the general dependence of the correction function on the optical depth for a set of selected helium lines. For this we calculated the correction function for $0 \leq \tau \leq 100$, $n_e=100$ cm$^{-3}$ and $T = 10000$ K and 20000 K. The dependence of correction functions on optical depth $\tau$ for different helium lines is presented on Fig. \ref{fig:odf1}. The Figure is directly comparable with Fig. 2 from \cite{bss2002}. On Fig. \ref{fig:odf1} results of \cite{bss2002} are denoted with dashed lines, while our results are denoted with solid ones.

The dependence of the correction functions on optical depth for all lines exhibits a generally similar trend in both models, yet noticeable differences arise and increase with optical depth. For a more quantitative comparison, we plot the relative difference
\begin{equation}
    \delta(\lambda, \tau) = \frac{f_{BSS}(\lambda, \tau) - f_{new}(\lambda, \tau)}{ f_{BSS}(\lambda, \tau)}
\end{equation}

\begin{figure}[htbp]
    \centering
    \includegraphics[width=0.95\linewidth]{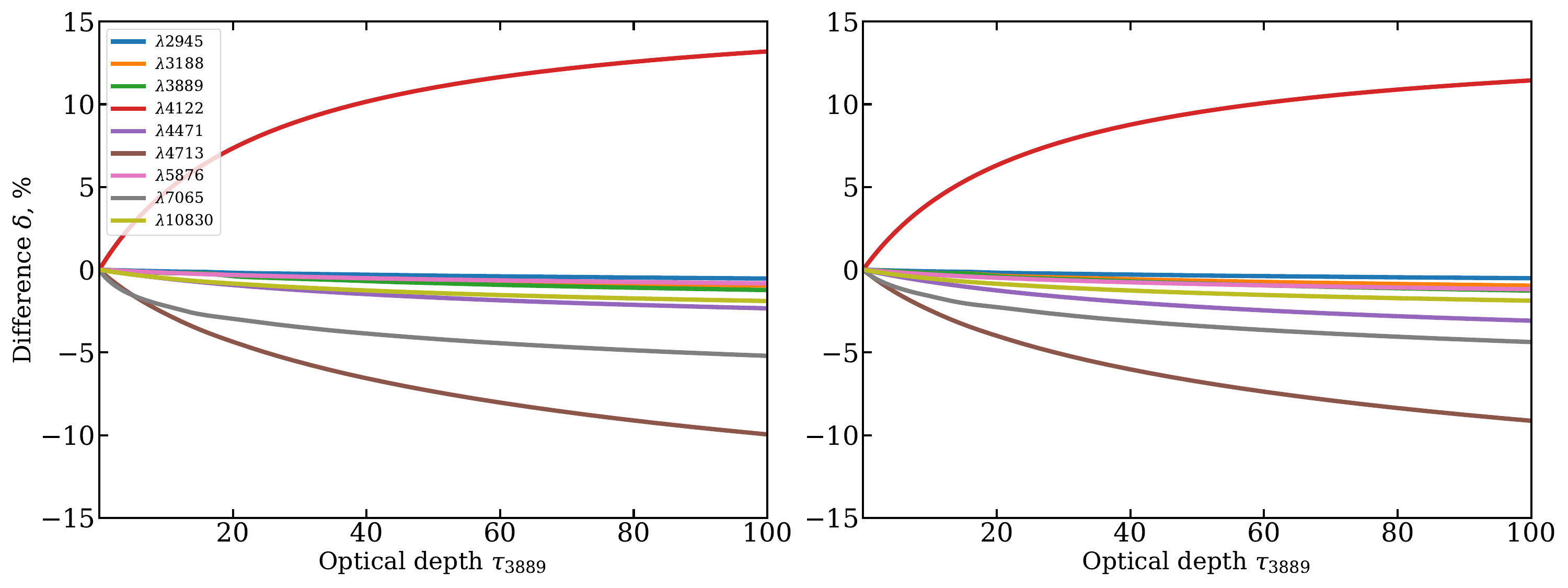}
    \caption{Relative difference $\delta(\lambda, \tau) = (f_{BSS}(\lambda, \tau) - f_{new}(\lambda, \tau)) / f_{BSS}(\lambda, \tau)$ in percent for cases with electron density $n_e=100$ cm$^{-3}$ and temperature $T = 10000$ K (left panel) and 20000 K (right panel).}
    \label{fig:odf2}
\end{figure}

The function $\delta(\lambda, \tau)$ is presented in Fig. \ref{fig:odf2}. The figure shows that the correction functions for lines originating from the $n^3P \rightarrow 2^3S$ transitions ($\lambda2945$, $\lambda3189$, $\lambda3889$, $\lambda10830$) remain essentially identical, with deviations $\leq 1\%$. In contrast, the correction functions for other lines (e.g., $\lambda4122$ and $\lambda7065$) differ by several percent even in the low optical depth region. This discrepancy is significant in the context of observational data analysis, since line fluxes are typically measured with an accuracy of only a few percent. The observed differences arise because our model is built upon a more advanced and accurate atomic data. 

It should also be noted that radiative transfer effects are significant primarily for triplet transitions. For singlet transitions, the influence of the relatively high population of the metastable level is negligible: the deviations of $f_\tau$ from unity are $\sim 10^{-3} - 10^{-4}$. These small deviations arise due to collisional mixing between the singlet and triplet ladders. Under the temperatures and densities considered here, this mixing remains weak, which explains the minimal departure from unity.

\begin{figure}[tbp]
    \centering
    \includegraphics[width=0.95\linewidth]{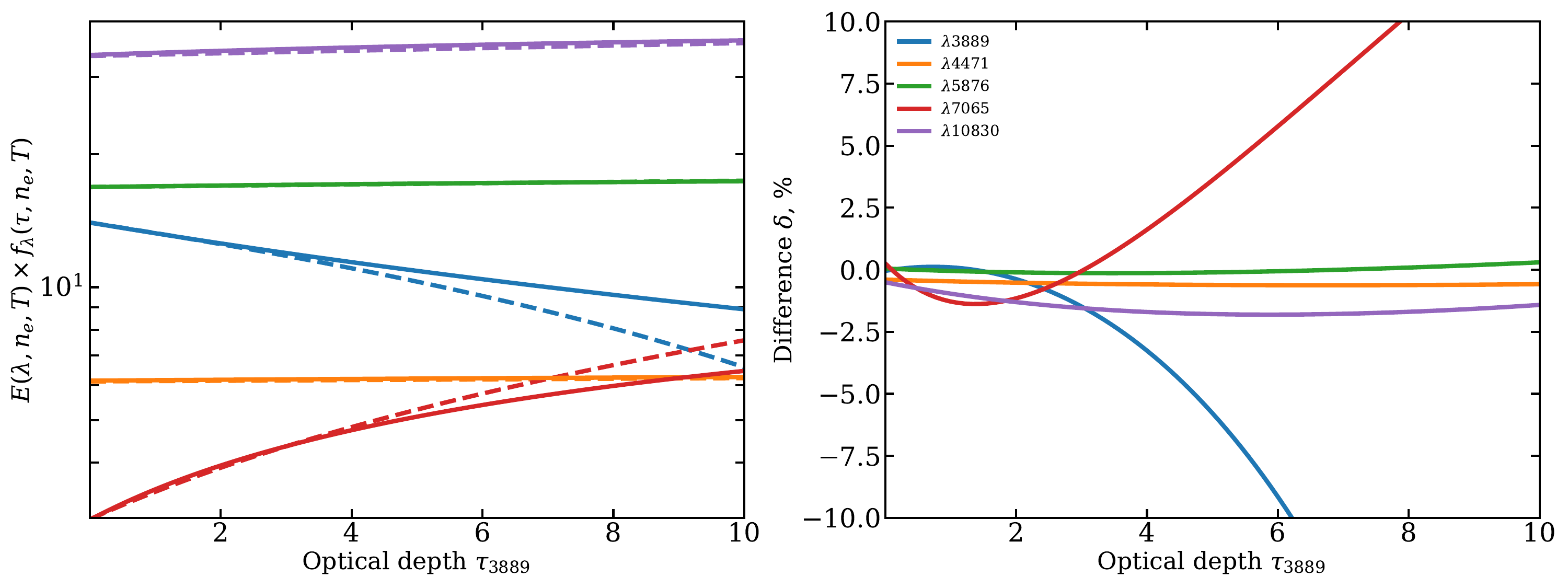}
    \caption{Emissivities of several He\,I lines with incorporated radiative transfer corrections as a function of optical depth. Emissivities are calculated for electron density $n_e = 100$ cm$^{-3}$ and temperature $T=10000$ K. Solid lines represent emissivities calculated using our model He\,I recombination spectrum. Dashed lines represent product $E(\lambda, n_e, T) \times f_{BSS}(\lambda, \tau, n_e, T)$ based on previous studies (see text). The left panel shows absolute values of emissivities in units $10^{26}$ erg cm$^3$ sec$^{-1}$. The right panel shows the percentage difference between them.}
    \label{fig:odf3}
\end{figure}

\begin{figure}[tbp]
    \centering
    \includegraphics[width=0.95\linewidth]{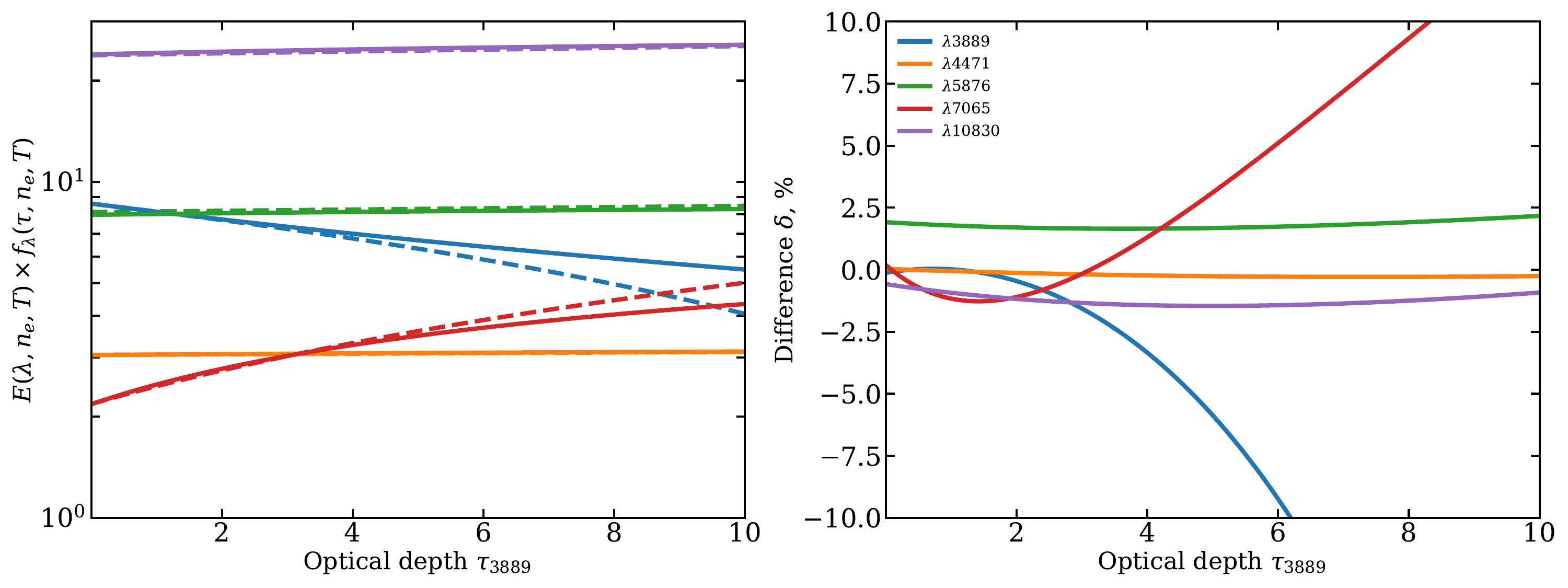}
    \caption{Same as Fig.~\ref{fig:odf3}, but emissivities are calculated for electron density $n_e = 100$ cm$^{-3}$ and temperature $T=20000$ K.}
    \label{fig:odf4}
\end{figure}

We compare the theoretical emissivities at a given optical depth, calculated with our model, against values expressed as $E(\lambda, n_e, T) \times f_{BSS}(\lambda, \tau, n_e, T)$. As is customary in all recent studies, the theoretical emissivities $E(\lambda, n_e, T)$ were adopted from \cite{aver2013} and optical depth function $f_{BSS}(\lambda, \tau, n_e, T)$ is taken from \cite{bss2002}. The results for $\tau \leq 10$, $n_e = 100$ cm$^{-3}$, and $T = 10000$ K and $20000$ K are shown in Fig.~\ref{fig:odf3} and \ref{fig:odf4}.

Fig.~\ref{fig:odf3} and \ref{fig:odf4} illustrates the deviations between the He\,I line emissivities obtained in the present study and those predicted by the commonly used approximation $E(\lambda, n_e, T) \times f_{BSS}(\lambda, \tau, n_e, T)$ as a function of optical depth. For most transitions, the discrepancies remain within 1–2\%, but for the $\lambda3889$ and $\lambda7065$ lines discrepancies are substantially larger. Specifically, usage of the correction function from \cite{bss2002} leads to systematic underestimation of the emissivity of $\lambda3889$ (by up $\sim$20\% or more with increasing optical depth) and overestimation of the emissivity of $\lambda7065$ by a comparable amount. Even within the physically realistic range of $\tau \lesssim 5$, the disagreement persists at the 5–10\% level, which is significant for diagnostic applications of these lines.  

Such pronounced discrepancies in $\lambda3889$ and $\lambda7065$ arise primarily from the linear dependence on optical depth adopted in  correction function from \cite{bss2002}. Clearly, this dependence is valid only in the limit of small optical depths ($\tau \lesssim 1$), which is not always satisfied in realistic conditions. This combines with the previously discussed deviations of order 1–2\% caused by the use of more advanced atomic data compound the effect. Taken together, these factors lead to a substantial overall disagreement, which must be carefully considered in future studies and in the analysis of helium emission lines.

\subsection{\label{subsec:approx}Approximation of the correction function}

For practical use in line diagnostics, calculation of optical depth correction function using full interpolation over the 3D parameter grid of $n_e, T$ and $\tau$ is somewhat inconvenient. This is due to performance issues that arise during the analyses of observational data, since the parameter grids are quite extensive. A reasonable way out in this case is to obtain approximation formulas, similar to the ones from \cite{bss2002}, which would be based on the results of our updated model. 

We derive approximation formulas for the following He\,I lines: $\lambda2945$, $\lambda3188$, $\lambda3889$, $\lambda4026$, $\lambda4471$, $\lambda4713$, $\lambda58760$, $\lambda7065$ and $\lambda10830$. The choice of lines is governed by these lines being the strongest triplet lines of He\,I in near UV, visible and near IR ranges. As noted previously, for singlet lines correction function equals unity with the accuracy of $\sim 10^{-3}-10^{-4}$. We approximate optical depth correction function $f_\tau(n_e, T)$ using the following relation:
\begin{equation}
    f_\tau(n_e, T) = \frac{1 + a(n_e, T) \times\tau}{1 + b \times\tau}
    \label{eq:fit_odf}
\end{equation}

Here parameter $a$ is a smooth function of electron density and temperature and $b$ is constant. This form of approximation function was chosen because (i) it exactly equals to 1 for $\tau =0$ and (ii) it depends on $\tau$ almost linearly in the region $\tau \lesssim  1$, i.e., demonstrates the same behavior as the approximation from \cite{bss2002}, while allowing for smooth non-linear behavior at higher optical depths. 

As shown previously, accounting for radiative transfer may lead to either a decrease or an increase in the emissivities of He\,I lines. A decreasing in emissivity with increasing optical depth is typical for the $n^3P \rightarrow 2^3S$ transitions, with the exception of the $\lambda10830$ line corresponding to $2^3P \rightarrow 2^3S$. In contrast, other lines from the triplet transition ladder exhibit enhanced emissivity as optical depth grows.

We fit the parameter $a$ as function of the electron density $n_e$ and temperature $T$ using the following relation:
\begin{equation}
a(n_e, T) = \sum_{i=0}^5 A_i(T) x^i~,~x=\lg\left(\frac{n_e}{10^2 \text{ cm}^{-3}}\right)
\end{equation}

Here fit parameters $A_i(T)$ are smooth functions of $T$ defined via the following equation:
\begin{equation}
    A_i(T) = \sum_{j=0}^3 B_i^{(j)} t^j~,~t=\lg\left(\frac{T}{10^4 \text{ K}}\right)
\end{equation}

Overall, the optical depth correction function described by eq. (\ref{eq:fit_odf}) has 25 fit parameter for each line: $b$, which determines the dependence on optical depth $\tau$, and 24 parameters  $B^{(i)}_j$ which govern the temperature and density dependence.

We calculate fit parameters for each line on 3D grid of $n_e$, $T$ and $\tau$ using trust reflective algorithm. Resulting fit parameters are summarized in Table \ref{tab:odf_fit_coefs}. The obtained approximation formulas reproduce the directly calculated correction function with an accuracy of $\lesssim\,0.1\%$ within the following parameter range: $1 \leq n_e \leq 10^4$ cm$^{-3}$, $8000 \leq T \leq 22000$ K, $0 \leq \tau \leq 10$.

\begin{table}[htbp]
\caption{\label{tab:odf_fit_coefs}
Fitting parameters $b$ and $B_j^{(i)}$ for optical depth correction function defined via eq. (\ref{eq:fit_odf}) for a set of He\,I triplet emission lines. Notation e$Y$ represents $\times 10^Y$.}
\centering
    \begin{tabular}{ccccccccc}
    \hline
& $B_0$ & $B_1$ & $B_2$ & $B_3$ & $B_0$ & $B_1$ & $B_2$ & $B_3$ \\
\hline
& \multicolumn{4}{c}{$\lambda 2945$, $b= $ 3.409e-02} & \multicolumn{4}{c}{$\lambda 4713$, $b= $ 4.276e-02} \\
\hline
$A_0$ & 4.241e-04 & 6.739e-05 & -1.184e-04 & 5.959e-05 & 7.655e-02 & -1.828e-02 & -1.008e-02 & 1.350e-02 \\
$A_1$ & 1.287e-04 & -1.141e-04 & 1.408e-05 & -1.418e-06 & -5.368e-04 & -6.202e-03 & -1.149e-02 & 2.845e-02 \\
$A_2$ & -9.949e-05 & 1.660e-04 & -9.626e-05 & -7.987e-05 & -9.923e-04 & -7.100e-03 & -1.264e-02 & 4.160e-02 \\
$A_3$ & -6.162e-05 & 3.444e-05 & -1.365e-04 & -1.267e-04 & -8.407e-04 & -2.651e-03 & 1.107e-02 & -2.577e-03 \\
$A_4$ & 5.217e-05 & -1.052e-04 & -4.027e-05 & 6.978e-06  & -6.045e-05 & 1.257e-03 & 1.165e-02 & -2.130e-02 \\
$A_5$ & -5.697e-06 & 2.270e-05 & 2.522e-05 & 4.756e-05 & 1.820e-04 & 3.486e-04 & -5.104e-03 & 5.408e-03 \\
\hline

& \multicolumn{4}{c}{$\lambda 3188$, $b= $ 5.131e-02} & \multicolumn{4}{c}{$\lambda 5876$, $b= $ 4.068e-02} \\
\hline
$A_0$ & 6.495e-04 & -5.338e-04 & 5.681e-04 & -2.681e-04  & 4.507e-02 & 3.753e-03 & 1.882e-03 & -1.899e-03 \\
$A_1$ & -6.749e-04 & 5.945e-04 & -2.085e-04 & 1.430e-05  & -2.264e-04 & -1.975e-04 & -1.024e-03 & -5.707e-03 \\
$A_2$ & 4.537e-04 & -7.174e-04 & 5.522e-04 & -2.628e-04 & 8.758e-05 & -3.356e-04 & -1.676e-03 & -5.033e-03 \\
$A_3$ & 2.958e-04 & -3.411e-04 & -1.921e-04 & 3.284e-04 & 9.324e-05 & -3.737e-04 & -1.951e-03 & 4.196e-03 \\
$A_4$ & -2.650e-04 & 3.628e-04 & -5.278e-04 & 4.049e-04 & -7.755e-05 & -8.592e-05 & -3.971e-04 & 4.138e-03 \\
$A_5$ & 3.416e-05 & -5.223e-05 & 1.909e-04 & -1.719e-04 & 1.304e-05 & 8.815e-05 & 5.352e-04 & -1.834e-03 \\
\hline

& \multicolumn{4}{c}{$\lambda 3889$, $b= $ 5.854e-02} & \multicolumn{4}{c}{$\lambda 7065$, $b= $ 5.461e-02} \\
\hline
$A_0$ & 7.399e-04 & 2.506e-04 & -2.762e-04 & 1.422e-04  & 2.361e-01 & -8.759e-02 & -5.182e-02 & 8.783e-02 \\
$A_1$ & 1.806e-04 & -3.057e-04 & -8.839e-05 & 1.853e-04  & -1.132e-02 & -6.569e-02 & -2.734e-02 & 1.896e-01 \\
$A_2$ & -9.546e-05 & 1.822e-05 & -3.319e-04 & 2.257e-04  & -1.440e-02 & -7.036e-02 & 5.927e-02 & 1.441e-01 \\
$A_3$ & -9.595e-05 & -3.434e-05 & -2.756e-04 & 4.107e-04  & -6.555e-03 & 6.632e-03 & 1.216e-01 & -1.965e-01 \\
$A_4$ & 4.975e-05 & -1.233e-04 & 2.953e-05 & 1.700e-04 & 1.941e-03 & 3.070e-02 & 1.702e-02 & -1.396e-01 \\
$A_5$ & -2.042e-06 & 4.591e-05 & 4.476e-05 & -1.138e-04 & 9.679e-04 & -8.075e-03 & -3.082e-02 & 7.805e-02 \\
\hline

& \multicolumn{4}{c}{$\lambda 4026$, $b= $ -9.988e-03} & \multicolumn{4}{c}{$\lambda 10830$, $b= $ 6.368e-02} \\
\hline
$A_0$ & -9.095e-03 & 7.274e-04 & 3.047e-04 & -7.155e-05  & 7.667e-02 & -5.558e-03 & -6.043e-03 & 1.144e-02 \\
$A_1$ & 1.007e-04 & -7.632e-05 & 9.027e-05 & -4.993e-04  & -5.435e-03 & -1.455e-02 & 1.262e-02 & 3.525e-03 \\
$A_2$ & 3.797e-06 & -5.135e-06 & 1.064e-04 & -6.808e-04  & -3.630e-03 & 1.509e-03 & 3.039e-02 & -3.820e-02 \\
$A_3$ & -3.394e-05 & 1.106e-05 & -1.811e-04 & -3.631e-04  & 1.018e-03 & 7.582e-03 & -8.689e-03 & -3.769e-03 \\
$A_4$ & 1.486e-05 & -2.175e-05 & -1.611e-04 & 1.065e-04  & 1.281e-03 & -4.800e-04 & -1.722e-02 & 2.311e-02 \\
$A_5$ & -1.713e-06 & 4.692e-06 & 6.341e-05 & 7.705e-05  & -4.396e-04 & -9.710e-04 & 6.709e-03 & -6.684e-03 \\
\hline

& \multicolumn{4}{c}{$\lambda 4471$, $b= $ 2.118e-03} \\
\hline
$A_0$ & 4.204e-03 & 9.874e-04 & 8.248e-04 & -3.394e-04 \\
$A_1$ & -5.567e-04 & 4.938e-04 & -4.527e-04 & -8.866e-04 \\
$A_2$ & 4.332e-05 & -4.700e-05 & 8.129e-05 & -1.352e-03 \\
$A_3$ & 1.922e-04 & -2.217e-04 & -6.444e-05 & -4.032e-04 \\
$A_4$ & -8.242e-05 & 5.455e-05 & -2.957e-04 & 4.713e-04 \\
$A_5$ & 7.268e-06 & 5.672e-06 & 9.692e-05 & -2.058e-06 \\
\hline
    \end{tabular}
\end{table}

\section{\label{sec:conclusions}Conclusions}

In the paper He\,I line emissivities and the optical depth correction functions for a selected set of the strongest emission lines in the UV, optical, and IR ranges are calculated. The results are obtained using new collisional-radiative model of He\,I recombination spectrum, which is based on modern and accurate atomic data.

The key distinction of the present He\,I spectrum model, compared to earlier ones, lies in the explicit treatment of non-unity escape probability of photons emerging from transitions to the metastable $2^3S$ level. Efficient absorption of photons at the metastable level with a non-negligible population leads to a noticeable redistribution of the populations of other levels in the helium atom. Explicit accounting for this effect allows us to incorporate radiative transfer effects directly into the calculated emissivities.

A new set of He\,I emissivities for the strongest emission lines was calculated for the following electron density and temperature ranges: $1 \leq n_e \leq 10^4$ cm$^-3$ and $8000 \leq T \leq 22000$ K. The calculated emissivities of He\,I emission lines in the optically thin case show very good agreement with previous studies. Minor deviations $(\lesssim 0.5 \%)$ are explained by the treatment of collisional rates in earlier models, where linear interpolation over a pre-computed grid of thermally averaged collision strengths (restricted to transitions up to $n=4$) was used. In contrast, in the present work the averaged collision strengths are obtained by direct integration for a larger set of transitions, extending up to $n=5$. 

Using the new model of He\,I recombination spectrum, a correction function to account for radiative transfer effects in He\,I lines is computed. Compared to the function of \cite{bss2002}, obtained with a simplified model, the new function shows systematic deviations of $\sim$1–2\% within a realistic range of optical depths. These deviations arise from the use of a more advanced model based on modern and more accurate atomic data.

Noticeably larger deviations (up to $\sim$10\% within a realistic parameter range) are found for the product $E(\lambda, n_e, T) \times f_{\tau}(\lambda, n_e, T)$, which is widely used to describe the emissivity of He\,I lines at non-zero optical depth. Such significant discrepancies are due to three main reasons:

\begin{enumerate}
\item the new model is based on more recent atomic data and includes a larger number of levels;
\item the optical-depth function from \cite{bss2002} relies on a linear dependence on $\tau$, which breaks down for $\tau \gtrsim 1$ for some important lines;
\item in current photoionization models, the product $E(\lambda, n_e, T) \times f_{\tau}(\lambda, n_e, T)$ inadvertently mixes atomic datasets: $E(\lambda, n_e, T)$ is computed with modern atomic data \cite{porter2012}, while $f_{\tau}(\lambda, n_e, T)$ still relies on outdated ones \cite{bss2002}.
\end{enumerate}

To overcome the limited applicability of the function from \cite{bss2002} and achieve better consistency between the correction function and theoretical He\,I emissivities, we propose a new fitting function $f_{\mathrm{new}}(\lambda, \tau, n_e, T)$. The function smoothly covers the following ranges of parameters: electron density $1 \leq n_e \leq 10^4$ cm$^-3$, temperature $8000 \leq T \leq 22000$ K and optical depth $0 \leq \tau \leq 10$. The fitting function has the accuracy of $\lesssim 0.1$ \% within the specified parameter domain.

The obtained results can be easily implemented in all modern studies of the Primordial Helium abundance as well as in other studies concerned with the analysis of He\,I emission lines. 

\section*{Acknowledgments}
This work was supported by the Russian Science Foundation under grant no. 23-12-00166. O.A. Kurichin expresses deep gratitude to Dr. Francisco Guzm{\'a}n for his detailed and valuable explanation of the complexities of the physical picture of collisional transitions in atoms. The authors express a great appreciation to Prof. Igor Bray and Prof. Yuri Ralchenko for providing the tabulated collision cross sections and valuable advice on the topic.

\section*{Data availability}
Full table of theoretical He\,I emissivities on the specified grid of electron densities and temperatures is available in Zenodo repository \cite{zenodo}.

\bibliographystyle{unsrt}  
\bibliography{test_bib}  

\providecommand{\noopsort}[1]{}\providecommand{\singleletter}[1]{#1}%
\begin{thebibliography}{10}

\bibitem{he_1}
Y.~I. {Izotov}, T.~X. {Thuan}, and N.~G. {Guseva}.
\newblock {A new determination of the primordial He abundance using the He I {\ensuremath{\lambda}}10830 {\r{A}} emission line: cosmological implications}.
\newblock {\em \mnras}, 445(1):778--793, November 2014.

\bibitem{he_2}
Erik {Aver}, Keith~A. {Olive}, and Evan~D. {Skillman}.
\newblock {The effects of He I {\ensuremath{\lambda}}10830 on helium abundance determinations}.
\newblock {\em \jcap}, 2015(7):011--011, July 2015.

\bibitem{he_3}
Mabel {Valerdi}, Antonio {Peimbert}, Manuel {Peimbert}, and Andr{\'e}s {Sixtos}.
\newblock {Determination of the Primordial Helium Abundance Based on NGC 346, an H II Region of the Small Magellanic Cloud}.
\newblock {\em \apj}, 876(2):98, May 2019.

\bibitem{he_4}
Vital {Fern{\'a}ndez}, Elena {Terlevich}, Angeles~I. {D{\'\i}az}, and Roberto {Terlevich}.
\newblock {A Bayesian direct method implementation to fit emission line spectra: application to the primordial He abundance determination}.
\newblock {\em \mnras}, 487(3):3221--3238, August 2019.

\bibitem{he_5}
O.~A. {Kurichin}, P.~A. {Kislitsyn}, V.~V. {Klimenko}, S.~A. {Balashev}, and A.~V. {Ivanchik}.
\newblock {A new determination of the primordial helium abundance using the analyses of H II region spectra from SDSS}.
\newblock {\em \mnras}, 502(2):3045--3056, April 2021.

\bibitem{he_6}
Tiffany {Hsyu}, Ryan~J. {Cooke}, J.~Xavier {Prochaska}, and Michael {Bolte}.
\newblock {The PHLEK Survey: A New Determination of the Primordial Helium Abundance}.
\newblock {\em \apj}, 896(1):77, June 2020.

\bibitem{he_7}
Mabel {Valerdi}, Antonio {Peimbert}, and Manuel {Peimbert}.
\newblock {Chemical abundances in seven metal-poor H II regions and a determination of the primordial helium abundance}.
\newblock {\em \mnras}, 505(3):3624--3634, August 2021.

\bibitem{he_8}
Erik {Aver}, Danielle~A. {Berg}, Keith~A. {Olive}, Richard~W. {Pogge}, John~J. {Salzer}, and Evan~D. {Skillman}.
\newblock {Improving helium abundance determinations with Leo P as a case study}.
\newblock {\em \jcap}, 2021(3):027, March 2021.

\bibitem{he_9}
Erik {Aver}, Danielle~A. {Berg}, Alec~S. {Hirschauer}, Keith~A. {Olive}, Richard~W. {Pogge}, Noah S.~J. {Rogers}, John~J. {Salzer}, and Evan~D. {Skillman}.
\newblock {A comprehensive chemical abundance analysis of the extremely metal poor Leoncino Dwarf galaxy (AGC 198691)}.
\newblock {\em \mnras}, 510(1):373--382, February 2022.

\bibitem{he_10}
O.~A. {Kurichin}, P.~A. {Kislitsyn}, and A.~V. {Ivanchik}.
\newblock {Determination of H II Region Metallicity in the Context of Estimating the Primordial Helium Abundance}.
\newblock {\em Astron. Lett.}, 47(10):674--685, October 2021.

\bibitem{he_11}
Akinori {Matsumoto}, Masami {Ouchi}, Kimihiko {Nakajima}, Masahiro {Kawasaki}, Kai {Murai}, Kentaro {Motohara}, Yuichi {Harikane}, Yoshiaki {Ono}, Kosuke {Kushibiki}, Shuhei {Koyama}, Shohei {Aoyama}, Masahiro {Konishi}, Hidenori {Takahashi}, Yuki {Isobe}, Hiroya {Umeda}, Yuma {Sugahara}, Masato {Onodera}, Kentaro {Nagamine}, Haruka {Kusakabe}, Yutaka {Hirai}, Takashi~J. {Moriya}, Takatoshi {Shibuya}, Yutaka {Komiyama}, Keita {Fukushima}, Seiji {Fujimoto}, Takashi {Hattori}, Kohei {Hayashi}, Akio~K. {Inoue}, Shotaro {Kikuchihara}, Takashi {Kojima}, Yusei {Koyama}, Chien-Hsiu {Lee}, Ken {Mawatari}, Takashi {Miyata}, Tohru {Nagao}, Shinobu {Ozaki}, Michael {Rauch}, Tomoki {Saito}, Akihiro {Suzuki}, Tsutomu~T. {Takeuchi}, Masayuki {Umemura}, Yi~{Xu}, Kiyoto {Yabe}, Yechi {Zhang}, and Yuzuru {Yoshii}.
\newblock {EMPRESS. VIII. A New Determination of Primordial He Abundance with Extremely Metal-poor Galaxies: A Suggestion of the Lepton Asymmetry and Implications for the Hubble Tension}.
\newblock {\em \apj}, 941(2):167, December 2022.

\bibitem{he_12}
Hiroto {Yanagisawa}, Masami {Ouchi}, Akinori {Matsumoto}, Masahiro {Kawasaki}, Kai {Murai}, Kimihiko {Nakajima}, Kazunori {Kohri}, Yuma {Sugahara}, Kentaro {Nagamine}, Ichi {Tanaka}, Ji~Hoon {Kim}, Yoshiaki {Ono}, Minami {Nakane}, Keita {Fukushima}, Yuichi {Harikane}, Yutaka {Hirai}, Yuki {Isobe}, Haruka {Kusakabe}, Masato {Onodera}, Michael {Rauch}, and Hidenobu {Yajima}.
\newblock {EMPRESS. XV. A New Determination of the Primordial Helium Abundance Suggesting a Moderately Low $Y_\mathrm{P}$ Value}.
\newblock {\em arXiv e-prints}, page arXiv:2506.24050, June 2025.

\bibitem{bss2002}
Robert~A. {Benjamin}, Evan~D. {Skillman}, and Derck~P. {Smits}.
\newblock {Radiative Transfer Effects in He I Emission Lines}.
\newblock {\em \apj}, 569(1):288--294, April 2002.

\bibitem{lp2001}
Grzegorz \L{}ach and Krzysztof Pachucki.
\newblock Forbidden transitions in the helium atom.
\newblock {\em Phys. Rev. A}, 64:042510, Sep 2001.

\bibitem{bss1999}
Robert~A. {Benjamin}, Evan~D. {Skillman}, and Derck~P. {Smits}.
\newblock {Improving Predictions for Helium Emission Lines}.
\newblock {\em \apj}, 514(1):307--324, March 1999.

\bibitem{smiths1991}
Derck~P. {Smits}.
\newblock {Low-temperature recombination coefficients and nova-shell abundances.}
\newblock {\em \mnras}, 248:193, January 1991.

\bibitem{smiths1996}
Derck~P. {Smits}.
\newblock {Theoretical HeI line intensities in low-density plasmas}.
\newblock {\em \mnras}, 278(3):683--687, February 1996.

\bibitem{sawey1993}
P.M.J. Sawey and K.A. Berrington.
\newblock Collision strengths from a 29-state r-matrix calculation on electron excitation in helium.
\newblock {\em At. Data Nucl. Data Tables}, 55(1):81--142, 1993.

\bibitem{porter2012}
R.~L. {Porter}, G.~J. {Ferland}, P.~J. {Storey}, and M.~J. {Detisch}.
\newblock {Improved He I emissivities in the case B approximation}.
\newblock {\em \mnras}, 425(1):L28--L31, September 2012.

\bibitem{delzanna2022}
G.~{Del Zanna} and P.~J. {Storey}.
\newblock {Helium line emissivities for nebular astrophysics}.
\newblock {\em \mnras}, 513(1):1198--1209, June 2022.

\bibitem{porter2005}
R.~L. {Porter}, R.~P. {Bauman}, G.~J. {Ferland}, and K.~B. {MacAdam}.
\newblock {Theoretical He I Emissivities in the Case B Approximation}.
\newblock {\em \apj}, 622(1):L73--L75, March 2005.

\bibitem{aver2013}
Erik {Aver}, Keith~A. {Olive}, R.~L. {Porter}, and Evan~D. {Skillman}.
\newblock {The primordial helium abundance from updated emissivities}.
\newblock {\em \jcap}, 2013(11):017, November 2013.

\bibitem{berg2025}
Danielle~A. {Berg}, Ryan~L. {Sanders}, Alice~E. {Shapley}, Michael~W. {Topping}, Naveen~A. {Reddy}, Evan~D. {Skillman}, Erik {Aver}, Fergus {Cullen}, Callum~T. {Donnan}, James~S. {Dunlop}, Tucker {Jones}, Ali~Ahmad {Khostovan}, Derek~J. {McLeod}, Desika {Narayanan}, Pascal~A. {Oesch}, Anthony~J. {Pahl}, Max {Pettini}, N.~M. {F{\"o}rster Schreiber}, and Daniel~P. {Stark}.
\newblock {The AURORA Survey: Robust Helium Abundances at High Redshift Reveal A Subpopulation of Helium-Enhanced Galaxies in the Early Universe}.
\newblock {\em arXiv e-prints}, page arXiv:2507.17057, July 2025.

\bibitem{bray2000}
I.~{Bray}, A.~{Burgess}, D.~V. {Fursa}, and J.~A. {Tully}.
\newblock {He (1 $^{1}$S, 2 $^{3}$S, 2 $^{1}$S, 2 $^{3}$P -> n $^{1,3}$ L): Thermally averaged electron collision strengths for n <= 5}.
\newblock {\em Astro. Astrophys. Suppl. Ser.}, 146:481--498, November 2000.

\bibitem{pdg2024}
S.~Navas et~al.
\newblock {Review of particle physics}.
\newblock {\em Phys. Rev. D}, 110(3):030001, 2024.

\bibitem{dm2007}
G.~W.~F. {Drake} and Donald~C. {Morton}.
\newblock {A Multiplet Table for Neutral Helium ($^{4}$He I) with Transition Rates}.
\newblock {\em \apjs}, 170(1):251--260, May 2007.

\bibitem{drake_handbook}
Gordon W.~F. {Drake}.
\newblock {\em {Springer Handbook of Atomic, Molecular, and Optical Physics}}.
\newblock Springer Nature, 2006.

\bibitem{drake1991}
G.~W.~F. Drake and R.~A. Swainson.
\newblock Quantum defects and the 1/n dependence of rydberg energies: Second-order polarization effects.
\newblock {\em Phys. Rev. A}, 44:5448--5459, Nov 1991.

\bibitem{hs1998}
D.~G. {Hummer} and P.~J. {Storey}.
\newblock {Recombination of helium-like ions - I. Photoionization cross-sections and total recombination and cooling coefficients for atomic helium}.
\newblock {\em \mnras}, 297(4):1073--1078, July 1998.

\bibitem{bates1949}
D.~R. {Bates} and Agnete {Damgaard}.
\newblock {The Calculation of the Absolute Strengths of Spectral Lines}.
\newblock {\em Philos. Trans. R. Soc. A}, 242(842):101--122, July 1949.

\bibitem{hey2017}
J~Hey.
\newblock On forms of the coulomb approximation as a useful source of atomic data for the spectroscopy of astrophysical and fusion plasmas.
\newblock {\em J. Phys. B}, 50:065701, 03 2017.

\bibitem{klarsfeld1989}
S.~Klarsfeld.
\newblock Alternative forms of the coulomb approximation for bound-bound multipole transitions.
\newblock {\em Phys. Rev. A}, 39:2324--2332, Mar 1989.

\bibitem{regem1979}
H~Regemorter, Hoang Dy, and M~Prud'homme.
\newblock Radial transition integrals involving low or high effective quantum numbers in the coulomb approximation.
\newblock {\em J. Phys. B}, 12:1053, 01 2001.

\bibitem{sobelman}
Igor~I. {Sobelman}.
\newblock {\em {Atomic spectra and radiative transitions}}.
\newblock Springer-Verlag, 1979.

\bibitem{varsh}
D.~A. {Varshalovich}, A.~N. {Moskalev}, and V.~K. {Khersonskii}.
\newblock {\em {Quantum Theory of Angular Momentum}}.
\newblock World Scientific Publishing Company, 1988.

\bibitem{draine}
Bruce~T. {Draine}.
\newblock {\em {Physics of the Interstellar and Intergalactic Medium}}.
\newblock Princeton University Press, 2011.

\bibitem{sh1991}
P.J. Storey and D.G. Hummer.
\newblock Fast computer evaluation of radiative properties of hydrogenic systems.
\newblock {\em Comput. Phys. Commun.}, 66(1):129--141, 1991.

\bibitem{drake1986}
G.~W.~F. Drake.
\newblock Spontaneous two-photon decay rates in hydrogenlike and heliumlike ions.
\newblock {\em Phys. Rev. A}, 34:2871--2880, Oct 1986.

\bibitem{bauman2005}
R.~P. {Bauman}, R.~L. {Porter}, G.~J. {Ferland}, and K.~B. {MacAdam}.
\newblock {J-Resolved He I Emission Predictions in the Low-Density Limit}.
\newblock {\em \apj}, 628(1):541--554, July 2005.

\bibitem{osterbrock}
Donald~E. {Osterbrock} and Gary~J. {Ferland}.
\newblock {\em {Astrophysics of gaseous nebulae and active galactic nuclei}}.
\newblock University Science Books, 2006.

\bibitem{burgess1992}
A.~{Burgess} and J.~A. {Tully}.
\newblock {On the analysis of collision strengths and rate coefficients.}
\newblock {\em Astron. Astrophys.}, 254:436--453, February 1992.

\bibitem{bray1994}
Igor Bray.
\newblock Convergent close-coupling method for the calculation of electron scattering on hydrogenlike targets.
\newblock {\em Phys. Rev. A}, 49:1066--1082, Feb 1994.

\bibitem{fursa1995}
Dmitry~V. Fursa and Igor Bray.
\newblock Calculation of electron-helium scattering.
\newblock {\em Phys. Rev. A}, 52:1279--1297, Aug 1995.

\bibitem{delzanna2020}
G.~{Del Zanna}, P.~J. {Storey}, N.~R. {Badnell}, and V.~{Andretta}.
\newblock {Helium Line Emissivities in the Solar Corona}.
\newblock {\em \apj}, 898(1):72, July 2020.

\bibitem{ralchenko2008}
Yu. {Ralchenko}, R.~K. {Janev}, T.~{Kato}, D.~V. {Fursa}, I.~{Bray}, and F.~J. {de Heer}.
\newblock {Electron-impact excitation and ionization cross sections for ground state and excited helium atoms}.
\newblock {\em At. Data and Nuc. Data Tables}, 94(4):603--622, July 2008.

\bibitem{ps1964}
R.~M. {Pengelly} and M.~J. {Seaton}.
\newblock {Recombination spectra, II}.
\newblock {\em \mnras}, 127:165, January 1964.

\bibitem{vos2001}
D.~Vrinceanu and M.~R. Flannery.
\newblock Classical and quantal collisional stark mixing at ultralow energies.
\newblock {\em Phys. Rev. A}, 63:032701, Feb 2001.

\bibitem{guzman1}
F.~{Guzm{\'a}n}, N.~R. {Badnell}, R.~J.~R. {Williams}, P.~A.~M. {van Hoof}, M.~{Chatzikos}, and G.~J. {Ferland}.
\newblock {H, He-like recombination spectra - I. l-changing collisions for hydrogen}.
\newblock {\em \mnras}, 459(4):3498--3504, July 2016.

\bibitem{guzman2}
F.~{Guzm{\'a}n}, N.~R. {Badnell}, R.~J.~R. {Williams}, P.~A.~M. {van Hoof}, M.~{Chatzikos}, and G.~J. {Ferland}.
\newblock {H-, He-like recombination spectra - II. l-changing collisions for He Rydberg states}.
\newblock {\em \mnras}, 464(1):312--320, January 2017.

\bibitem{guzman4}
N.~R. {Badnell}, F.~{Guzm{\'a}n}, S.~{Brodie}, R.~J.~R. {Williams}, P.~A.~M. {van Hoof}, M.~{Chatzikos}, and G.~J. {Ferland}.
\newblock {H, He-like recombination spectra - IV. Clarification and refinement of methodology for l-changing collisions}.
\newblock {\em \mnras}, 507(2):2922--2929, October 2021.

\bibitem{guzman3}
F.~{Guzm{\'a}n}, M.~{Chatzikos}, P.~A.~M. {van Hoof}, Dana~S. {Balser}, M.~{Dehghanian}, N.~R. {Badnell}, and G.~J. {Ferland}.
\newblock {H-, He-like recombination spectra - III. n-changing collisions in highly excited Rydberg states and their impact on the radio, IR, and optical recombination lines}.
\newblock {\em \mnras}, 486(1):1003--1018, June 2019.

\bibitem{lmixing}
D.~{Vrinceanu}, R.~{Onofrio}, J.~B.~R. {Oonk}, P.~{Salas}, and H.~R. {Sadeghpour}.
\newblock {Efficient Computation of Collisional {\ensuremath{\ell}}-mixing Rate Coefficients in Astrophysical Plasmas}.
\newblock {\em \apj}, 879(2):115, July 2019.

\bibitem{vos2012}
D.~{Vrinceanu}, R.~{Onofrio}, and H.~R. {Sadeghpour}.
\newblock {Angular Momentum Changing Transitions in Proton-Rydberg Hydrogen Atom Collisions}.
\newblock {\em \apj}, 747(1):56, March 2012.

\bibitem{vos2017}
D.~{Vrinceanu}, R.~{Onofrio}, and H.~R. {Sadeghpour}.
\newblock {On the treatment of {\ensuremath{\ell}}-changing proton-hydrogen Rydberg atom collisions}.
\newblock {\em \mnras}, 471(3):3051--3056, November 2017.

\bibitem{Deliporanidou}
E.~{Deliporanidou}, N.~R. {Badnell}, P.~J. {Storey}, G.~{Del Zanna}, and G.~J. {Ferland}.
\newblock {H, He-like recombination spectra VI: quadrupole l-changing collisions}.
\newblock {\em \mnras}, 539(4):2957--2966, June 2025.

\bibitem{lebedev1998}
Vladimir~S. {Lebedev} and Israel~L. {Beigman}.
\newblock {\em {Physics of Highly Excited Atoms and Ions}}, volume~22.
\newblock Springer, Berlin, Heidelberg, 1998.

\bibitem{guzman5}
F.~{Guzm{\'a}n}, M.~{Chatzikos}, and G.~J. {Ferland}.
\newblock {H-, He-like recombination spectra {\textendash} V. On the dependence of the simulated line intensities on the number of electronic levels of the atoms}.
\newblock {\em \mnras}, 539(4):2939--2956, June 2025.

\bibitem{seaton1959}
M.~J. {Seaton}.
\newblock {Radiative recombination of hydrogenic ions}.
\newblock {\em \mnras}, 119:81, January 1959.

\bibitem{zenodo}
O.~A. {Kurichin} and A.~V. {Ivanchik}.
\newblock {He I Emissivity Table}.
\newblock Zenodo Repository, 2025.
\newblock 10.5281/zenodo.17195750.

\end{thebibliography}

\newpage
\appendix

\section{\label{app:a_val_coef}Fitting coefficients for A-values}

Fitting coefficients for the oscillator strengths of different transitions of the form $(n, l, s) ~\longrightarrow (n', l', s)$. The values of quantum numbers for the lower-level are denoted with a prime and are considered fixed. The Einstein coefficient for an allowed transition is determined by the formula:
\begin{equation}
    A_{nl\rightarrow n'l'} = \frac{0.667 \text{ cm$^2$ sec$^{-1}$}}{\lambda^2} \frac{g_{n'l'}}{g_{nl}} f_{nl\rightarrow n'l'}
\end{equation}

Oscillator strength is calculated via equation:
\begin{equation}
    f_{nl\rightarrow n'l'} = \frac{1}{\nu^3}~exp\left ( a\,x^2 + b\,x + c\right)
\end{equation}

Here $\nu$ is the effective quantum number of the $(n, l, s)$ level, which is calculated using quantum defects, $x = ln(E_{n'l's} / \Delta E)$, where $\Delta E = E_{nls} - E_{n'l's}$. Fitting coefficients $a, b, c$ are given in Table \ref{tab:a_val_coefs}.

\begin{table}[ht]
    \centering
    \caption{\label{tab:a_val_coefs} Fitting coefficients $a, b, c$ used to calculate A-values for dipole allowed transitions in He atom via the extrapolation of transition data from \cite{dm2007}.}
    \centering
    \begin{tabular}{cccccccc}
    \hline
    Transition     &  $a$ & $b$ & $c$ & Transition     &  $a$ & $b$ & $c$\\
    \hline
    $n^1P ~\rightarrow ~ 2^1S$ & -0.5212 & 1.4980 & 0.8258 & $n^3P ~\rightarrow ~ 2^3S$ & -1.2552 & 0.9575 & 0.3055  \\
    $n^1S ~\rightarrow ~ 2^1P$ & -0.0667 & 2.3183 & -1.4292 & $n^3S ~\rightarrow ~ 2^3P$ & -0.0062 & 2.3165 & -1.3656  \\
    $n^1D ~\rightarrow ~ 2^1P$ & -0.2041 & 3.1697 & 1.1341 & $n^3D ~\rightarrow ~ 2^3P$ & -0.1696 & 2.8455 & 1.3102 \\
    \hline
    $n^1P ~\rightarrow ~ 3^1S$ & -0.3629 & 1.6108 & 1.2713 & $n^3P ~\rightarrow ~ 3^3S$ & -0.8665 & 1.2341 & 0.6799 \\
    $n^1S ~\rightarrow ~ 3^1P$ & -0.0569 & 2.3246 & -0.3603 & $n^3S ~\rightarrow ~ 3^3P$ & -0.0115 & 2.3183 & -0.2894\\
    $n^1D ~\rightarrow ~ 3^1P$ & -0.2200 & 2.4906 & 1.7920 & $n^3D ~\rightarrow ~ 3^3P$ & -0.2697 & 2.2448 & 1.8427 \\
    $n^1P ~\rightarrow ~ 3^1D$ & -0.3660 & 2.9286 & -2.6734 & $n^3P ~\rightarrow ~ 3^3D$ & -0.2400 & 2.7619 & -1.9216 \\
    $n^1F ~\rightarrow ~ 3^1D$ & -0.1381 & 3.0896 & 1.2853 & $n^3F ~\rightarrow ~ 3^3D$ & -0.1943 & 3.3380 & 1.4267\\
    \hline
    $n^1P ~\rightarrow ~ 4^1S$ & -0.3006 & 1.7246 & 1.6071 & $n^3P ~\rightarrow ~ 4^3S$ &-0.7341 & 1.4638 & 0.9610\\
    $n^1S ~\rightarrow ~ 4^1P$ & -0.0467 & 2.3228 & 0.3040 & $n^3S ~\rightarrow ~ 4^3P$ & -0.0078 & 2.3139 & 0.3696\\
    $n^1D ~\rightarrow ~ 4^1P$ & -0.1468 & 2.2748 & 2.1984 & $n^3D ~\rightarrow ~ 4^3P$ & -0.1985 & 2.0782 & 2.1801\\
    $n^1P ~\rightarrow ~ 4^1D$ & -0.2738 & 2.7738 & -1.4383 & $n^3P ~\rightarrow ~ 4^3D$ & -0.1814 & 2.6539 & -0.7813\\
    $n^1F ~\rightarrow ~ 4^1D$ & -0.2268 & 2.5973 & 1.9472 & $n^3F ~\rightarrow ~ 4^3D$ & -0.2575 & 2.7239 & 2.0930 \\
    $n^1D ~\rightarrow ~ 4^1F$ & -0.3916 & 3.3412 & -3.3231 & $n^3D ~\rightarrow ~ 4^3F$ & -0.3843 & 3.3271 & -2.9957\\
    $n^1G ~\rightarrow ~ 4^1F$ & -0.2711 & 3.7808 & 1.5045 & $n^3G ~\rightarrow ~ 4^3F$ & -0.2721 & 3.7828 & 1.5314\\
    \hline
    $n^1P ~\rightarrow ~ 5^1S$ & -0.2758 & 1.8294 & 1.8733 & $n^3P ~\rightarrow ~ 5^3S$ & -0.6850 & 1.6723 & 1.1762 \\
    $n^1S ~\rightarrow ~ 5^1P$ & -0.0408 & 2.3223 & 0.7859 & $n^3S ~\rightarrow ~ 5^3P$ & -0.0054 & 2.3102 & 0.8460\\
    $n^1D ~\rightarrow ~ 5^1P$ & -0.1065 & 2.1902 & 2.5024 & $n^3D ~\rightarrow ~ 5^3P$ & -0.1575 & 2.0307 & 2.4379\\
    $n^1P ~\rightarrow ~ 5^1D$ & -0.2243 & 2.6910 & -0.6640 & $n^3P ~\rightarrow ~ 5^3D$ & -0.1469 & 2.5900 & -0.0655\\
    $n^1F ~\rightarrow ~ 5^1D$ & -0.1675 & 2.3629  & 2.3354 & $n^3F ~\rightarrow ~ 5^3D$ & -0.1887 & 2.4414 & 2.4813\\
    $n^1D ~\rightarrow ~ 5^1F$ & -0.3040 & 3.1131 & -1.9805 & $n^3D ~\rightarrow ~ 5^3F$ & -0.3021 & 3.1079  & -1.7250  \\
    $n^1G ~\rightarrow ~ 5^1F$ & -0.3177  & 3.1087  & 2.2473 & $n^3G ~\rightarrow ~ 5^3F$ & -0.3182 & 3.1100  & 2.2861 \\
    $n^1F ~\rightarrow ~ 5^1G$ & -0.4604 & 3.8089  & -3.4727  & $n^3F ~\rightarrow ~ 5^3G$ & 0.0 & 2.9204 & -3.0172 \\
    $n^1H ~\rightarrow ~ 5^1G$ & -0.3193 & 4.1129 & 1.4451 & $n^3H ~\rightarrow ~ 5^3G$ & -0.3180 & 4.1104 & 1.4591 \\
    \hline
    $n^1P ~\rightarrow ~ 6^1S$ & -0.2647 & 1.9245 & 2.0906 & $n^3P ~\rightarrow ~ 6^3S$ & -0.6805 & 1.8866 & 1.3347 \\
    $n^1S ~\rightarrow ~ 6^1P$ & -0.0359 & 2.3203 & 1.1646 & $n^3S ~\rightarrow ~ 6^3P$ &-0.0014 & 2.3024 & 1.2210 \\
    $n^1D ~\rightarrow ~ 6^1P$ & -0.0887 & 2.1630 & 2.7455 & $n^3D ~\rightarrow ~ 6^3P$ & -0.1439 & 2.0427 & 2.6420\\
    $n^1P ~\rightarrow ~ 6^1D$ & -0.1928 & 2.6388  & -0.1011  & $n^3P ~\rightarrow ~ 6^3D$ & -0.1262  & 2.5514 & 0.4557\\
    $n^1F ~\rightarrow ~ 6^1D$ & -0.1287 & 2.2530  & 2.6123 & $n^3F ~\rightarrow ~ 6^3D$ & -0.1398 & 2.2988 & 2.7609  \\
    $n^1D ~\rightarrow ~ 6^1F$ & -0.2540  & 2.9795  & -1.1413 & $n^3D ~\rightarrow ~ 6^3F$ & -0.2505  & 2.9720 & -0.9195 \\
    $n^1G ~\rightarrow ~ 6^1F$ & -0.2374 & 2.7265 & 2.6690  & $n^3G ~\rightarrow ~ 6^3F$ & -0.2382  & 2.7281 & 2.7173  \\
    $n^1F ~\rightarrow ~ 6^1G$ & -0.3637  & 3.4985  & -2.1657 & $n^3F ~\rightarrow ~ 6^3G$ & -0.3553  & 3.4705  & -2.0873 \\
    $n^1H ~\rightarrow ~ 6^1G$ & -0.3608  & 3.4253 & 2.3087  & $n^3H ~\rightarrow ~ 6^3G$ & -0.3609  & 3.4253  & 2.3218 \\
    $n^1G ~\rightarrow ~ 6^1H$ & -0.4846  & 4.1284  & -3.8739  & $n^3G ~\rightarrow ~ 6^3H$ & -0.4846 & 4.1384  & -3.8739 \\
    $n^1I ~\rightarrow ~ 6^1H$ & -0.3591  & 4.4223 & 1.2858  & $n^3I ~\rightarrow ~ 6^3H$ & -0.3607 & 4.4249 & 1.2940 \\
    \hline
    $n^1P ~\rightarrow ~ 7^1S$ & -0.2675 & 2.0248 & 2.2657 & $n^3P ~\rightarrow ~ 7^3S$ & -0.7162 & 2.1488 & 1.4243 \\
    $n^1S ~\rightarrow ~ 7^1P$ & -0.0359 & 2.3203  & 1.1646 & $n^3S ~\rightarrow ~ 7^3P$ & 0.0007 & 2.2974 & 1.5292 \\
    $n^1D ~\rightarrow ~ 7^1P$ & -0.0765 & 2.1523  & 2.9516 & $n^3D ~\rightarrow ~ 7^3P$ & 0.0 & 1.7561 & 2.9838  \\
    $n^1P ~\rightarrow ~ 7^1D$ & -0.1745 & 2.6098 & 0.3368 & $n^3P ~\rightarrow ~ 7^3D$ & -0.1105  & 2.5211  & 0.8670 \\
    $n^1F ~\rightarrow ~ 7^1D$ & -0.1002 & 2.1876 & 2.8356   & $n^3F ~\rightarrow ~ 7^3D$ & -0.1111 & 2.2258  & 2.9811 \\
    $n^1D ~\rightarrow ~ 7^1F$ & -0.2189 & 2.8875 & -0.5330  & $n^3D ~\rightarrow ~ 7^3F$ & 0.0 & 2.3661 & -0.0380 \\
    $n^1G ~\rightarrow ~ 7^1F$ & -0.1744 & 2.5022 & 2.9645 & $n^3G ~\rightarrow ~ 7^3F$ & -0.1744 & 2.5023 & 3.0202 \\
    $n^1F ~\rightarrow ~ 7^1G$ & -0.2992 & 3.2931 & -1.3247 & $n^3F ~\rightarrow ~ 7^3G$ & -0.2958 & 3.2783 & -1.2494 \\
    $n^1H ~\rightarrow ~ 7^1G$ & -0.2744 & 2.9744 & 2.8118 & $n^3H ~\rightarrow ~ 7^3G$ & -0.2722 & 2.9693 & 2.8277 \\
    $n^1G ~\rightarrow ~ 7^1H$ & -0.3931 & 3.7817 & -2.5038 & $n^3G ~\rightarrow ~ 7^3H$ & 0.0 & 2.8458 & -1.9605 \\
    $n^1I ~\rightarrow ~ 7^1H$ & -0.3976 & 3.7215 & 2.2533 & $n^3I ~\rightarrow ~ 7^3H$ & -0.3976 & 3.7215 & 2.2533  \\
    $n^1H ~\rightarrow ~ 7^1I$ & -0.5194 & 4.4687 & -4.3716 & $n^3H ~\rightarrow ~ 7^3I$ & -0.5194 & 4.4749 & -4.3716 \\
    \hline
    \end{tabular}
\end{table}

\section{\label{app:rrr_coef}Fitting coefficients for scaling factor}

Fitting coefficients for scaling factor $f(T, \gamma)$, which is used to calculate radiative recombination rate to sublevel $\gamma = [n, l, s]$:
\begin{equation}
    \alpha_\gamma(T) = f(T, \gamma) \times \alpha_{\rm H}(T, n, l)
\end{equation}

Scaling factor $f(T, \gamma)$ is defined via the following equation:
\begin{equation}
    f(T, \gamma) = \frac{a_1(T)}{n^{a_2(T)}} + a_3(T)
    \label{eq:rrr_scale_eq_2}
\end{equation}

Here $a_i(T)$ are fit coefficients. Their dependence on temperature is determined by a fourth-degree polynomial with temperature $T$ in K with fitting coefficients $b_j^{(i)}$ being presented in Table \ref{tab:rr_scale_fit_coefs}:
\begin{equation}
    a_i(T) = \sum_{j=0}^4 b_j^{(i)} \left( \frac{T}{10^4~\text{K}}\right)^j
\end{equation}

\begin{table}[ht]
\caption{\label{tab:rr_scale_fit_coefs}
Fitting coefficients used to calculate scaling term for radiative recombination rate to sublevels in He atom with $l\leq2$ and $s=1, 3$. Notation e$Y$ represents $\times 10^Y$.}
\centering
    \begin{tabular}{cccccc}
    \hline
    & $b_4$ & $b_3$ & $b_2$ & $b_1$ & $b_0$ \\
\hline
\multicolumn{6}{c}{$l = 0$, $s = 1$} \\
\hline
$a_1$ & 3.84e-03 & -2.75e-02  & 7.72e-02  & -9.76e-02  & 1.72e-01 \\
$a_2$ & 1.34e-02 & -9.85e-02  & 2.89e-01  & -4.09e-01  & 1.24e+00 \\
$a_3$ & -6.69e-04 & 5.34e-03  & -1.86e-02  & 4.71e-02  & 1.28e-01 \\
\hline
\multicolumn{6}{c}{$l = 0$, $s = 3$} \\
\hline
$a_1$ & 1.62e-02 & -1.16e-01  & 3.21e-01  & -3.53e-01  & 6.98e-01 \\
$a_2$ & 1.65e-02 & -1.22e-01  & 3.56e-01  & -4.96e-01  & 1.36e+00 \\
$a_3$ & -1.98e-03 & 1.60e-02  & -5.67e-02  & 1.56e-01  & 1.89e-01 \\
\hline
\multicolumn{6}{c}{$l = 1$, $s = 1$} \\
\hline
$a_1$ & -1.95e-03 & 1.30e-02  & -3.48e-02  & 4.73e-02  & -9.39e-02 \\
$a_2$ & 8.52e-03 & -4.67e-02  & 8.96e-02  & -5.46e-02  & 1.26e+00 \\
$a_3$ & 2.69e-04 & -2.13e-03  & 6.96e-03  & -1.38e-02  & 2.77e-01 \\
\hline
\multicolumn{6}{c}{$l = 1$, $s = 3$} \\
\hline
$a_1$ & 3.36e-03 & -2.87e-02  & 9.56e-02  & -1.48e-01  & 5.49e-01 \\
$a_2$ & -1.95e-04 & -5.29e-03  & 3.62e-02  & -8.47e-02  & 1.16e+00 \\
$a_3$ & -1.65e-03 & 1.32e-02  & -4.62e-02  & 1.18e-01  & 7.41e-01 \\
\hline
\multicolumn{6}{c}{$l = 2$, $s = 1$} \\
\hline
$a_1$ & 1.49e-03 & -9.16e-03  & 2.17e-02  & -2.57e-02  & 2.57e-02 \\
$a_2$ & 8.49e-02 & -4.94e-01  & 1.07e+00  & -1.03e+00  & 1.47e+00 \\
$a_3$ & 1.99e-05 & -1.26e-05  & -4.23e-04  & 1.44e-03  & 2.47e-01 \\
\hline
\multicolumn{6}{c}{$l = 2$, $s = 3$} \\
\hline
$a_1$ & 2.30e-03 & -1.24e-02  & 2.40e-02  & -2.12e-02  & -4.22e-02 \\
$a_2$ & -6.50e-02 & 4.08e-01  & -1.00e+00  & 1.30e+00  & 2.49e-01 \\
$a_3$ & 8.27e-04 & -5.62e-03  & 1.47e-02  & -1.72e-02  & 7.99e-01 \\
\hline
    \end{tabular}
\end{table}

\end{document}